\pdfoutput=1
\documentclass[12pt]{article}

\usepackage{multirow}
\usepackage{mathptmx}
\usepackage{graphicx}
\usepackage{times}
\usepackage{amssymb}
\usepackage{amsmath}

\usepackage{amssymb}
\usepackage{latexsym}

\usepackage{url}
\usepackage{xcolor}

\definecolor{newcolor}{rgb}{.8,.349,.1}

\usepackage{hyperref}
\usepackage{amsmath}
\usepackage{ dsfont }
\usepackage{xspace}
\usepackage{booktabs}
\usepackage{multirow}
\usepackage{subcaption}
\usepackage{siunitx}
\usepackage{etoolbox}
\usepackage{rotating}

\newcommand{\Task}[1]{\emph{Task #1}\xspace}
\newcommand{\task}{\Task}
\newcommand{\mse}{\emph{MSE}\xspace}
\newcommand{\accuracy}{\emph{accuracy}\xspace}
\newcommand{\percent}{\emph{percent10}\xspace}
\newcommand{\deltametric}{\emph{delta10}\xspace}
\newcommand{\md}{\texttt{AM}\xspace}
\newcommand{\nl}{\texttt{NL}\xspace}
\newcommand{\vgg}{\texttt{VGG16}\xspace}
\newcommand{\lenet}{\texttt{LeNet}\xspace}
\newcommand{\MD}{\md}
\newcommand{\NL}{\nl}
\newcommand{\comment}[1]{}

\newcommand{\ubold}{\fontseries{b}\selectfont}
\robustify\ubold%
\newcommand{\best}{\ubold}

\newcommand{\nbgraphs}{\num{27000} }
\title{Toward automatic comparison of visualization techniques: application to graph visualization}

\author{L. Giovannangeli, R. Bourqui, R. Giot and D. Auber \\
{\small \{loann.giovannangeli, romain.bourqui, romain.giot, david.auber\}@labri.fr}}
\date{March, 2020}

\begin{document}
\maketitle

\begin{abstract}
Many end-user evaluations of data visualization techniques have been run during the last decades. Their results are cornerstones to build efficient visualization systems. However, designing such an evaluation is always complex and time-consuming and may end in a lack of statistical evidence and reproducibility. We believe that modern and efficient computer vision techniques, such as deep convolutional neural networks (CNNs), may help visualization researchers to build and/or adjust their evaluation hypothesis. 
The basis of our idea is to train machine learning models on several visualization techniques to solve a specific task. Our assumption is that it is possible to compare the efficiency of visualization techniques based on the performance of their corresponding model. 
As current machine learning models are not able to strictly reflect human capabilities, including their imperfections, such results should be interpreted with caution. 
However, we think that using machine learning-based pre-evaluation, 
as a pre-process of standard user evaluations, should help researchers to perform a more exhaustive study of their design space.
Thus, it should improve their final user evaluation by providing it better test cases.
In this paper, we present the results of two experiments we have conducted to assess how correlated the performance of users and computer vision techniques can be.
That study compares two mainstream graph visualization techniques: node-link (\NL) and adjacency-matrix (\MD) diagrams.
Using two well-known deep convolutional neural networks, we partially reproduced user evaluations from Ghoniem \textit{et al.} and from Okoe \textit{et al.}. These experiments showed that some user evaluation results can be reproduced automatically. 
\end{abstract}

\section{Introduction}
\label{sec_introduction}

Information visualization has now been established as a fruitful strategy in various application domains for exploration of large and/or complex data. When designing a new \emph{visualization technique}, it is necessary to \emph{assess its efficiency} compared to existing ones. While computational performances can be easily evaluated,  
it is more complicated to assess user performances. This is usually achieved by carrying out a \emph{user evaluation} that compares several techniques on one or several \emph{tasks} (\textit{e.g.} \cite{ghoniem2004comparison,ghoniem2005readability,okoe2017revisited, okoe2018node}). During such a study, a population of users is requested to solve several times the same task(s) using various input data presented with different visualization techniques or different variations of a technique. Experimental results (\textit{e.g.}, accuracy or completion time) are then statistically validated to confirm or refute some preliminary hypothesis. Completing a user evaluation can become time-consuming as the number of techniques, the considered dataset, and the number of tasks increase. As that may bias the experimental results due to loss of attention or tiredness, it is necessary to maintain the completion time of the entire evaluation reasonable. To that end, one usually tries to keep the number of \textit{experimental trials}  reasonable~\cite{purchase2012experimental} by reducing the number of evaluated techniques, the number of tasks and/or the dataset. This is one of the strongest bottlenecks of user evaluations as visualization designers have to reduce the explored design space to a few visualization techniques without any strong evidence. 

We believe that using automatic comparison of visualization techniques, as a pre-evaluation step, should help to address this issue. 
As no user is involved in such an evaluation, visualization designers could assess the relative efficiency of several visualization techniques, or variations of one.
A formal user evaluation would then be carried out based on the results of that pre-evaluation.

Deep neural networks systems~\cite{lecun2015deep} are used in more and more scenarios (\emph{e.g}, natural language processing, instance segmentation, image captioning) and outperform well-established state-of-the-art methods
since the publication of the AlexNet in 2012~\cite{krizhevsky2012imagenet}, and its great performance in the ImageNet challenge~\cite{russakovsky2015imagenet}.
 \cite{behrisch2018quality} mention, in their survey on quality metrics in visualization, that using supervised machine learning approaches, such as deep learning, is a promising direction for evaluating the quality of a visualization. \cite{haleem2018evaluating} made use of such systems in an information visualization context and showed that these techniques can be widely beneficial to our community. 
 
This paper presents the results of a study conducted to assess how correlated the performance of users and computer vision techniques are.
The basis of that study consists in: first, generate a set of still images for each compared visualization technique from a common dataset, then, train models for each images-set and task. 
Consider two visualization techniques $A$ and $B$, a model architecture $M$ and the bests-trained models $M_A$ and $M_B$ respectively on $A$ and $B$ for a given task. 
The study aims at answering the following question: if $M_A$ performs more accurately than $M_B$, can we observe the same performance difference with a population of users ? 
Computer vision algorithms and users performances have already been compared~\cite{stablinger2016, dodge2018}; sometimes, machine learning algorithms provide better results, especially when the model architecture focuses on a specific task~\cite{He2015surpassingHuman} while users outperform when bias (\emph{e.g.} distortion, noise) is added to the images~\cite{dodge2017}.
However, none of these studies compared machine learning models with users in a context of information visualization.
Our study reproduces four tasks from two previous user evaluations~\cite{ghoniem2004comparison,ghoniem2005readability,okoe2017revisited, okoe2018node}.
It aims at comparing node-link diagram (\NL) and adjacency matrix diagrams (\MD) for three low-level tasks (\textit{counting}) and one task of higher level (\textit{minimum distance between two nodes}). That evaluation involves about \num{230000} experimental objects. 

The results that come through this study are drawing good indicators to this approach; however they must be considered within the scope of their limitations which are raised all along the experimental process.
 

The remainder of this paper is organized as follows. Section~\ref{sec_previous} presents related works with a focus on machine learning for counting, machine learning for information visualization and user evaluation of graph visualization techniques. Section~\ref{sec_methodology} describes the general approach proposed to automatically compare visualizations and its different steps. Section~\ref{sec_experimental_evaluation} describes the experimental evaluations made to assess the pertinence of the method and their results while section~\ref{sec_discussion} discusses them. Finally, Section~\ref{sec_conclusion} draws conclusions and presents directions for future works.

\section{Related works}
\label{sec_previous}
The proposed study mainly relates to the use of \emph{machine learning for information visualization}. 
As mentioned in section~\ref{sec_introduction}, we identified four tasks of \textit{counting} and \textit{connectivity}. 
While machine learning has already been used for \textit{counting}, there is, to the best of our knowledge, no prior work related to the shortest path between two nodes. 
We also present the results of \emph{user evaluations of graph visualizations}, emphasizing some inconsistencies, to justify the choice of tasks we used.

\subsection{Machine learning for information visualization}

\textbf{Graphical Perception} \cite{haehn2018evaluating} have reproduced~\cite{Cleveland} settings with four different neural networks. 
Their goal was to show how these networks compare to human. They show that CNNs perform as well as or better than humans for almost every task, although CNNs cannot estimate edge length ratios properly. In that work, MultiLayer Perceptron (MLP), LeNet~\cite{lecun1998gradient}, VGG~\cite{simonyan2014very} and Xception~\cite{chollet2017xception} networks have been tested. In each case, results were better if the networks were completely retrained specifically for a specific task. We use a similar set up for our experiment by completely retraining our CNNs.

\textbf{Readability of Force Directed Graph Layouts} \cite{haleem2018evaluating} work is the most similar in spirit to ours. They proposed to use handcrafted CNN (inspired by VGG) in order to find properties of force directed layouts in node link diagrams. 
While the focus of that research is to design an optimized CNN architecture for that particular task, our study relies on standard model architectures. 
We assume these models, designed by the machine learning community, are efficient~\cite{haehn2018evaluating}, 
and we use them to compare various visualization techniques.

\subsection{Machine learning for counting objects}
In this paper, we aim to compare several counting tasks and a connectivity one on two graph visualization techniques (see Section~\ref{sec_experimental_evaluation}) using machine learning. Recent advances in that domain have proven that such tasks may be automatically done but also provide some cues to set up our experimentation.

\textbf{Complete image versus patches} ~\cite{lempitsky2010learning} propose a semi-supervised learning framework for visual object counting tasks. They recover the density function of objects in an image thanks to a linear transformation from the feature space that represents each pixel. A tailored loss function helps to optimize the weights of this linear transformation while the integration of the density function provides the number of objects. The method has been validated on cell counting on fluorescent microscopy images and people counting in surveillance video. \cite{walach2016learning} improve the former work by 20\%  with an approach based on CNNs in a boosting framework (CNN$_{i+1}$ learns to estimate the residual error of CNN$_i$) and image patches. Of course, this method is not exact, and the quality of the results may vary according to the input images. Measuring these differences is the basis of our comparison approach. 
While that method improves the performances, we use full image search method in our experiment since we do not make any assumption on the size of the objects to be counted.  

\textbf{Objects identification} 
Recent research also proved that deep networks may learn automatically the kinds of graphical objects they needed to solve a task. \cite{segui2015learning} estimate the count (up to 5) of even digits in an image. They show that the network can extract features that encode the objects to count whereas they have never been explicitly provided to the network. Another result is the one of \cite{rahnemoonfar2017deep}; they used a deep learning framework to count fruits in images. The originality of this work resides in the use of artificial images during the training while still achieving more than 90\% of prediction on 100 real images of tomato plants. Therefore, for graph counting tasks, it may be possible to use the same learning technique on different visualizations without specific changes according to the way entities and relations are visually encoded. 
This will be the basis of the approach used in our study: to apply the same learning technique (model architecture, metrics, etc.) on different visualizations so that the visualization itself is the only parameter which should impact the performances of the training and trained models.

\subsection{User evaluation of graph visualization}
\label{sec:user_eval_graph}

In this paper, we focus on the evaluation of two of the most widely accepted graph visualization techniques: nodelink (\NL) and adjacency matrix (\MD) diagrams. 
They are state-of-the-art graph visualization techniques many researchers try to improve, for example by optimizing an aesthetic criterion or an objective function, or to identify some criteria of the visualization that can facilitate task's solving (\textit{e.g.} \cite{mueller2007comparison, kobourov2014crossings}).

Concurrently, user evaluations have also been performed to compare the efficiency of these two techniques~\cite{ghoniem2004comparison,ghoniem2005readability,keller2006matrices,abuthawabeh2013finding,alper2013weighted,okoe2015ecological,okoe2017revisited, okoe2018node}. 
 On one hand, \cite{ghoniem2004comparison,ghoniem2005readability} evaluated these techniques on seven \textit{counting} and \textit{connectivity} tasks and concluded that \NL performs better than \MD for connectivity tasks when graphs are small and sparse while \MD outperforms \NL in all the other conditions. These results were later reproduced by the user evaluation of~\cite{keller2006matrices} for similar tasks even with small graphs. In their evaluation,~\cite{alper2013weighted} compared two variations of \NL and \MD for dynamic weighted graphs and concluded that \MD performs better than \NL for \textit{weight variations} and \textit{connectivity} tasks. On the other hand,~\cite{abuthawabeh2013finding} did not find any difference in term of user performance when considering higher level tasks of the software engineering domain. \cite{sansen15} compared four visual edge encodings in \MD among which one corresponds to a \NL where nodes have been duplicated for the \textit{path finding} task. This evaluation showed very few significative differences between \MD and their variation of \NL. Such contradictory results were as well obtained by other researches~\cite{okoe2015ecological,okoe2017revisited,okoe2018node}. \cite{okoe2015ecological} partially reproduced two \textit{connectivity} tasks out the seven tasks of~\cite{ghoniem2004comparison,ghoniem2005readability} with some modifications of the experimental setup. These modifications consisted in using a realworld dataset, ordering the matrix to emphasize the cluster structures and allowing node drag in the \NL. Another important difference is the size of the involved population as running their experiment with an online system allowed them to recruit a larger number of participants. This evaluation's results showed that \NL outperforms \MD on each tested task. 
A larger scale study, presented by \cite{okoe2017revisited, okoe2018node}, explored a larger range of tasks as it included fourteen tasks classified into  \textit{topology/connectivity}, \textit{group} and \textit{memorability} tasks. That study showed that \NL outperforms \MD for \textit{topology/connectivity} and \textit{memorability} tasks while the contrary can be observed for \textit{group} tasks.
Finally, a recent study by \cite{ren2019NLvsAM} on \num{600} subjects showed that \NL performs better than \MD for \textit{adjacency, accessibility} and \textit{connectivity} tasks. The downside of this study is that it has been performed with only \num{2} graphs of \num{20} and \num{50} vertices, and relatively low densities ($\leq 0.1$); therefore its results must be compared carefully to previous studies.

\begin{figure*}[!ht]
\centering
\includegraphics[width=0.95\linewidth]{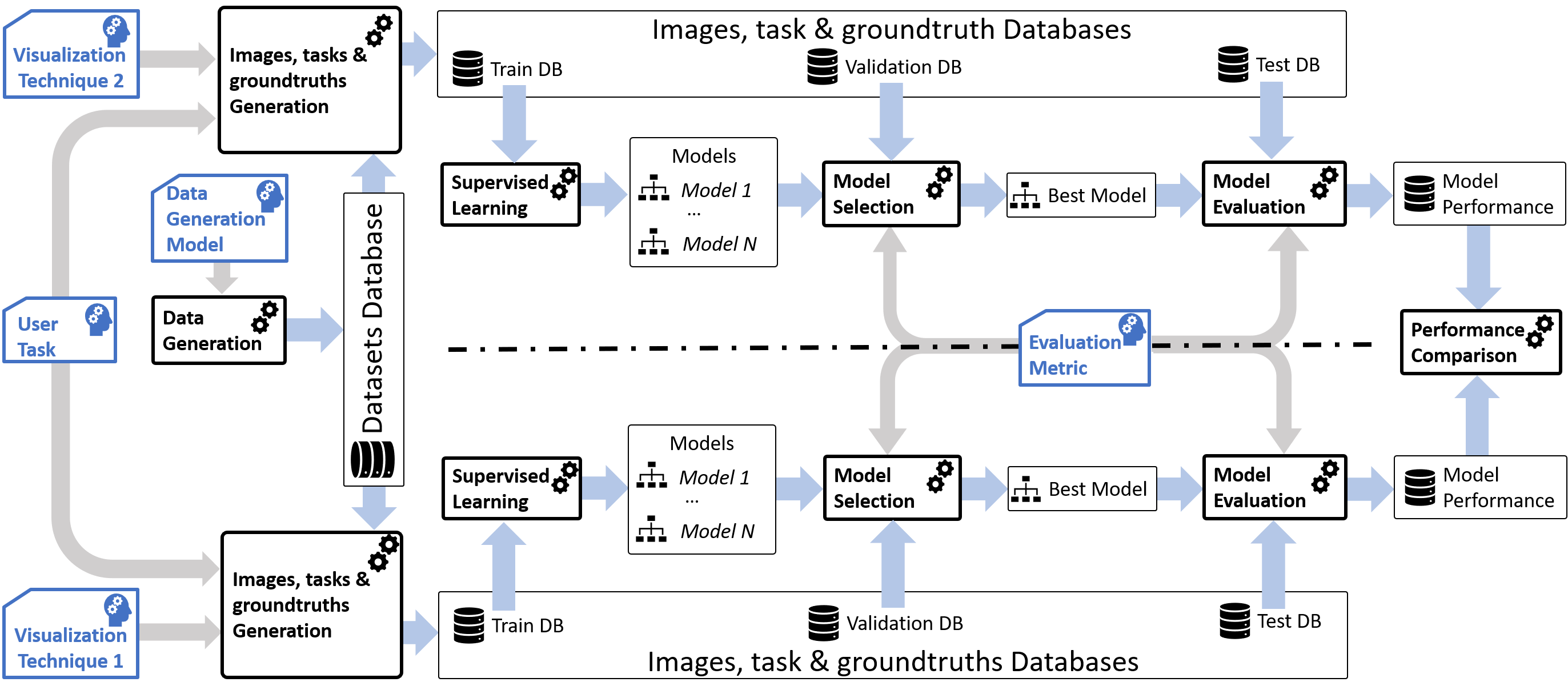}
\caption{Illustration of the experimental protocol proposed in this work.
First steps consist of generating data and their representations with the visualization techniques to compare.
Next steps consist of training models, selecting the best ones and evaluating them on unseen data.
Finally, the efficiencies of the visualization techniques are compared using usual statistical methods as for any quantitative user evaluation. Blue icons 
indicate where visualization experts are needed.}
\label{fig_methodology}
\end{figure*}

We believe that the divergence in these results are mainly due to three aspects. First of all, even though the evaluation protocols are similar, their slight differences (available interaction tools, wording of the questions) can drastically induce variations in the results. The second aspects relates to the populations involved in these studies. As their size (\textit{e.g.}, from few dozens~\cite{ghoniem2004comparison,ghoniem2005readability} to few hundred~\cite{okoe2017revisited, okoe2018node}) and demographics varied (\textit{e.g.}, post-graduate students and researchers in computer science~\cite{ghoniem2004comparison,ghoniem2005readability} and Amazon Mechanical Turk users~\cite{okoe2017revisited, okoe2018node}), the results can vary as well. The last aspect relates to the evaluation datasets as these evaluations are performed in distinct contexts and therefore use different datasets having various topological properties. Using an automated evaluation as a pre-process of the user evaluation may help to solve such reproducibility issues as far as the datasets, the model architectures and the visualization protocols are provided.
Considering these previous studies, we decided to run our experiment on four tasks having no contradictory result.
In order to be able to compare our results to those of existing user evaluations, we partially reproduced those of \cite{ghoniem2004comparison,ghoniem2005readability} and \cite{okoe2018node}. 
Three of our tasks are low-level ones and relate to \textit{counting}, while the last task, of higher level, consists in determining the length of the shortest between two highlighted nodes.

\section{Automatic evaluation with machine learning: general approach}
\label{sec_methodology}
The key idea of the proposed method is to use machine learning models to solve a specific task for each visualization technique to be compared. Then, the efficiency of these techniques can be evaluated by comparing the performance of their related models.
While bio-inspired models~\cite{Thomas2014} aim to mimic human behavior, many modern machine learning models do not perform like humans but rather try to perform better than them~\cite{He2015surpassingHuman}. If such a model cannot solve a task on a visualization, it could mean either that: (i) the model is too simple or not designed for the task; (ii) the task needs a high level of abstraction (\emph{e.g.} cultural knowledge) to be solved; or (iii) the visualization does not contain (enough) elements for the model to solve the task. 

Several important criteria for  \emph{Human Computer Interface} field may be estimated using such machine learning models.

\textbf{Learning convergence:} by comparing how fast, in term of number of epochs, a visualization can be understood by the model, one could estimate the learning curve to solve a task on this visualization. 

\textbf{Complexity of the task:} models have various structure complexity (\textit{e.g.} number of layers, type of operations in the layers, \textit{etc}.). By checking the performance of several well-known models, going from less to more complex, one could automatically evaluate the difficulty to solve a task on a visualization. Simple tasks are expected to work well with simple models while more complex tasks should work with more complex models. 

\textbf{Accuracy of a visualization:} by measuring the number of correct results for a task, one can compare a visualization to another. Errors are usually present when there is some ambiguity in the visualization due to, for instance, cluttering issues. Modern machine learning-based models perform quite well here, especially on counting tasks (\textit{e.g.} \cite{walach2016learning, rahnemoonfar2017deep}). Thus, one can assume that if the model cannot solve the ambiguity, it will also be hard or at least time consuming to be solved in a standard evaluation.

These criteria rely on the main assumption our study aims at validating: a correlation can be made between humans and machines performances. We are aware this assumption is not self-evident and may be dependent on the task type, its dataset, the model's architecture and other parameters. If this assumption is to be refuted, this approach can enable us to explore (dis)similarities between humans and machines performances.

Figure~\ref{fig_methodology} summarizes our experimental protocol while the remaining of the section presents each block of our evaluation process: the \emph{data generation}, the \emph{images, tasks and ground truths}, the \emph{model selection} and the \emph{model evaluation and Performance comparison}. It also presents the \emph{role} that remains to \emph{visualization experts}.

\subsection{Data generation}
Our evaluation protocol should be applied in a context for which we know which tasks have to be handled.
Consequently, we should have a set of raw input data (to be visualized with the different techniques) associated with the expected answer for each of these tasks. Such dataset can either be generated automatically using a data generation algorithm or collected from existing data source.  
Our experiment has shown that the data generation process requires significant attention. 
Indeed, as we detail in the next sections, data distribution is of utmost importance: the more uniform the distribution is, the easier the setup of evaluation and training will be. 
It is not always easy or feasible to generate datasets respecting this constraint.
One of our main will for this approach being the reproducibility of the evaluation, generated data can be stored and reused for future experimentation. It enables the community to create test beds to compare new visualization techniques with previous ones.

\subsection{Images, tasks and ground truths}

The second step consists of generating visual representations for each input data with each visualization technique. They will be used during the experiment. In our current approach, we only deal with static visualizations and do not support interactive ones.
Thus, that step consists of generating static labeled images on which a model or a user should answer a question (\textit{i.e.}, solve a task). 
We note $v_i = (img_i, grd_i)$ a data sample,  where $img_i$ is the image (corresponding to a visualization technique) and $grd_i$ is the ground truth to be predicted among the sequence of all possible answers to a specific task, noted $G$. 
 As previously stated, generating a set $V=\{v_1, \ldots , v_n\}$ of samples for a specific task is challenging as it has to follow a valid distribution. For instance, if the same value has to be predicted in $80\%$ of the cases and a model has an accuracy of $80\%$ for that task, it could mean that the model succeeded in detecting the bias in our experiment. 
 Ideally, the distribution should verify the following property:
\begin{equation}
\forall a, b \in G , 
\sum_{(img,grd)\in V} \mathds{1}\{a=grd\}
= 
\sum_{(img,grd)\in V} \mathds{1}\{b=grd\}
\label{eq:distrib}
\end{equation}
where $\mathds{1}\{a\} = \left\{ \begin{gathered} 1 \  \makebox{if} \ a\ \makebox{is}\ true \\ 0 \makebox{ otherwise} \end{gathered} \right.$.

In other words, each ground truth of $G$ should appear the same number of times in $V$ so its distribution is uniform.
Such experiment is usually not feasible with users since the number of experimental objects is too large making the completion time unrealistic.

The first step of a formal user evaluation is to let the user learn the visualization and the tasks with a training set. Our approach works similarly: $V$ is split into three sets called \emph{train} (to learn the model), \emph{validation} (to select the best model) and \emph{test} (to fairly evaluate the model) datasets; they are respectively noted $V_{train}$, $V_{validation}$ and $V_{test}$. Each of them should verify the property~\eqref{eq:distrib} and thus initial set $V$ must be large enough. 

\subsection{Model selection}
\label{sec:model_select}

The model selection step consists of choosing the model that performs the best on $V_{validation}$ on a given metric. 
It means that it has learned well with $V_{train}$ and is able to generalize with the dataset $V_{validation}$ (\emph{i.e.}, it does not suffer of overfitting). 
We apply standard machine learning methodology for that purpose.
With deep neural networks-based models, we train models during a fixed number of epochs ($100$ by default) and save their state as many times. 
Then we use the set $V_{validation}$ to compute the set of predictions for each epoch and save these predictions. For each epoch $e$ and each sample $v_i \in V_{validation}$, a prediction is the answer provided by the model and is noted $p_e(v_i) \in G$. We can then evaluate for each epoch its performance on a given metric regarding its predictions and the ground truths. 
In this experiment, we used four metrics:
\begin{itemize}
    \item \textbf{MSE (mean squared error)}: $mse(\mathbf{y}, \mathbf{\hat{y}}) = \sum_i^{Card\left(\mathbf{y}\right)} \frac{ |y_i - \hat{y}_i |^2  } {Card\left(\mathbf{y}\right)}$\\
    That is a common regression metric in the machine learning community that considers the labels proximity (lower is better).
    
    \item \textbf{Accuracy}: \begin{math}
    accuracy(\mathbf{y}, \mathbf{\hat{y}}) = \sum_{i=1}^{Card\left(\mathbf{y}\right)} \frac{ \mathds{1}\{y_i = \hat{y}_i\}  } {Card\left(\mathbf{y}\right)}
    \end{math}
    
    \item  \textbf{\percent}: $percent10(\mathbf{y}, \mathbf{\hat{y}}) = \sum_i^{Card\left(\mathbf{y}\right)} \frac{ \mathds{1} \left\{ .9 y_i <= \hat{y}_i <= 1.1 y_i \right\}  } {Card\left(\mathbf{y}\right)}$\\
    That metric is a variation of the \accuracy that allows +/-$10\%$ of errors (higher is better). 

    \item  \textbf{$\deltametric$}: $ delta10(\mathbf{y}, \mathbf{\hat{y}}) = \sum_i^{Card\left(\mathbf{y}\right)} \frac{ \mathds{1} \left\{ y_i - {10} <= \hat{y}_i <= y_i + {10} \right\}  } {Card\left(\mathbf{y}\right)}$ \\
    That metric is a variation of the \accuracy that allows +/-$10$ units of errors (higher is better).
\end{itemize}
where \begin{description}
\centering
\item $\mathbf{y} = \left[ grd\ |\ (img, grd) \in V_{validation}\right ] $, list of ground truths;
\item $\mathbf{\hat{y}} = \left[ p_e(v)\ |\ v \in V_{validation}\right ]$, list of associated predictions.
\end{description}
While \mse allows measuring the distance of the predictions to their associated ground truths, \percent and \deltametric consider predictions that are \textit{close} to their ground truths as correct.
Although \percent and  $\deltametric$ seem similar, they have opposite limitations. While \percent is more permissive when the predictions are associated with high-value ground truths than low-value ones, $\deltametric$ has the opposite behavior. Thus, both are sensitive to the ground truths average value and distribution.

\subsection{Model evaluation and Performance comparison}\label{sec:eval_metrics}
$V_{test}$ serves as input data to evaluate the selected model performances but also to ensure that it generalizes well. 
As the elements of $V_{test}$ have not been previously processed during the learning and model selection stages, we can measure the capacity of the model to solve a task on an unseen image.

For each $v \in V_{test}$, we compute $p_e(v)$ using the best epoch $e$ found during the model selection stage. We note $p_e(V_{test})$ the sequence of all predictions for the elements of $V_{test}$. 
When comparing $k$ visualization techniques (or variations), the $k$ corresponding models provide $k$ sequences of predictions to be compared. It is noteworthy to mention that the best epochs may be different for each combination of model's architecture, visualization technique, task and metric.

Then, we make sure the selected model has effectively learned how to solve the considered task. This is usually achieved by comparing the predictions of a model to random choices or computing evaluation metrics. In this work, we propose to use the \emph{coefficient of determination} ($R^2$) designed for regression problems.

\begin{center}
    \begin{math}
          R^2(\mathbf{y}, \mathbf{\hat{y}}) = 1 - {\sum\sum_{i=1}^{Card\left(\mathbf{y}\right)} (y_i - \hat{y}_i)^2} / {\sum_{i=1}^{Card\left(\mathbf{y}\right)} (y_i - \overline{y})^2}
    \end{math}
\end{center}
    \noindent where $\overline{y} = \sum_{i=1}^{Card\left(\mathbf{y}\right)} y_i / Card\left(\mathbf{y}\right)$
\noindent $1$ means that $\hat{y}$ explains perfectly $y$ (very good system), while 0 means the opposite. For our experiments, we'll consider that $R^2$ scores above \num{0.3} mean that the model learned to solve a task.

In case of a classification problem, we recommend the use of the Matthews correlation coefficient defined as:  
\begin{center}
    \begin{math}
        \begin{array}{l}
    MCC(\mathbf{y}, \mathbf{\hat{y}}) = \\\frac
        {\sum_{ijk} C_{ii}C_{jk} - C_{ij}C_{ki}}
        {
            \sqrt{\sum _i (\sum_j C_{ij})(\sum_{i\prime, j\prime | i\prime \neq i} C_{i\prime j\prime})} 
            \sqrt{\sum _i (\sum_j C_{ij})(\sum_{i\prime, j\prime | i\prime \neq i} C_{j\prime i\prime})}
        }
        \end{array}
    \end{math}
\end{center}
\noindent considering the confusion matrix $C$ (built from $\mathbf{\hat{y}}$ and $\mathbf{y}$) of a  $K$ classification problem. This metric's advantage is to be insensitive to unbalanced datasets.
A value of $1$ represents a perfect prediction, 0 a random one and $-1$ a total disagreement between the prediction its ground truth.

Finally, performance comparison consists of comparing the sequences of predictions of each model (\emph{i.e.} each visualization technique) with the ground truths. For that purpose, we use the previously presented metrics (see Section~\ref{sec:model_select}) to evaluate each prediction on $V_{test}$ and we use standard statistical tests (\emph{e.g.} Wilcoxon signed rank, Kruskal Wallis) to determine if the predictions of a model are significantly better than others.

\subsection{Role of visualization experts}

The very design of our evaluation protocol permits to run experiments without involving any human.
However visualization experts must remain in the process even in its early stages and play the same role as in formal user evaluations. 
In particular, the expert must set all the experimental parameters such as the visualization techniques to be compared, the considered tasks and hypothesis, the data generation model (or the collection of data) as well as the evaluation metrics (see Figure~\ref{fig_methodology}). \\

It is noteworthy to remind that our approach does not intend to replace humans evaluations.
It rather provides a tool to help experts (i) to choose the visualizations and their parameters to use during a user evaluation, and (ii) and to justify these choices with reproducible computed metrics.

\section{Experimental comparison of \NL and \MD}
\label{sec_experimental_evaluation}
This section presents two evaluations carried out to validate the proposed approach and its premise. We have chosen to reproduce user evaluation studies that serve as references to draw a first conclusion about its effectiveness. 
Thus, we reproduced 3 (low-level) tasks out of the 7 tasks of~\cite{ghoniem2004comparison,ghoniem2005readability} and 1 (higher-level) task of~\cite{okoe2018node}. 
These two experiments aim at comparing \NL and \MD representations. It is important to remind that, as in any other similar study of the literature, these experiments compare opinionated implementations of \NL and \MD views. 
For the sake of reading simplification, ``\NL" and ``\MD" use systematically refer to our implementation.

The following sections describe these two evaluations including their \emph{experimental protocol}, \emph{results} and a discussion about their \emph{main limitations}.

\subsection{Counting tasks of Ghoniem \textit{et al.}}
\label{GhoniemUseCase}

\subsubsection{Evaluation protocol}
\paragraph{\textbf{Task, visualization techniques and hypothesis}}
\label{sec_tasks_to_solve}

In this evaluation, we focus on the first three tasks of \cite{ghoniem2004comparison,ghoniem2005readability} that mainly relate to \textit{counting};
(i) \task{0}: exact count of the nodes number;
(ii) \task{1}: exact count of the edges  number;
(iii) \task{2}: maximum degree of the graph.
\Task{0} and \Task{1} are slight variations of Ghoniem \textit{et al.} first two tasks which consisted of providing \textit{estimates} of the number of nodes and edges. 
Giving an exact count being a harder problem than giving an approximation, we know our results will differ from the original study, in that the accuracies will most likely be lower. 
However, during both validation and evaluation, estimates of the ground-truths can be considered (see Section~\ref{sec:model_select}).
\Task{2} is a variation of Ghoniem \textit{et al.} third task as they considered identification of the node of maximum degree instead of the maximum degree. We decided to implement this task variation as we wanted to focus on \textit{counting} tasks (see Section~\ref{sec:user_eval_graph}). 
 
Several studies~\cite{ghoniem2004comparison, 	ghoniem2005readability, okoe2017revisited, okoe2018node,ren2019NLvsAM} have provided contradictory results about the efficiency of \NL and \MD on several tasks.
However, to the best of our knowledge, none of these results are related to the \num{3} ones we focus on. This was, to a certain extent, mandatory in this study as we expected our results to corroborate some commonly admitted ones to estimate our approach feasibility. 
Therefore, our hypothesis is:

\textit{H1:} \MD should outperform \NL for each of the three tasks.

\paragraph{\textbf{Graph dataset}} 
We generated our dataset with the same random graph model as in the original study but we could not use the same implementation of that model  as their generator is no more available.
Moreover, the number of graphs considered in our experiment is much larger.
On one hand, supervised learning methods require a large training dataset, and on the other hand, we have made the hypothesis that using many graphs would help to train models able to better generalize.
When generating the graphs of this study, we considered two parameters:
\begin{itemize} 
	\item the \textit{number of nodes} which varies from \num{20} to \num{100} (we refer to this parameter as the graph size).
	\item the \textit{edge density} as defined in~\cite{ghoniem2004comparison, ghoniem2005readability} which was set to \num{0.2}, \num{0.4} and \num{0.6} and corresponds to $\sqrt{\frac{2|E|}{|V|^2}}$.
\end{itemize}

There are $243$ combinations of parameters ($3 \makebox{ edge densities} * 81 \makebox{ graph size}$).
$\num{100}$ instances 
are generated for each combination, 
which led to $\num{24300}$ graphs for training and validation.
$\num{2700}$ additional graphs are generated for test while trying to follow the same distribution as for the training and validation datasets.
This leads to \nbgraphs graphs.

\paragraph{\textbf{Images datasets}}
\label{GhoniemImagesData}

\label{sec_representation}


\begin{figure*}[!tb]
	\centering

	 \begin{subfigure}{0.18\linewidth}
	    \includegraphics[width=\linewidth]{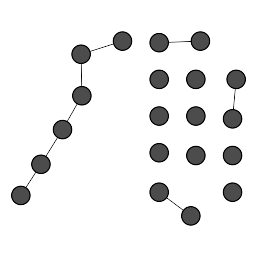}
		\caption{\emph{V} = 20, \emph{E} = 8, \emph{d} = 0.2}
	\end{subfigure}\hspace{1.5cm}
	 \begin{subfigure}{0.18\linewidth}
	    \includegraphics[width=\linewidth]{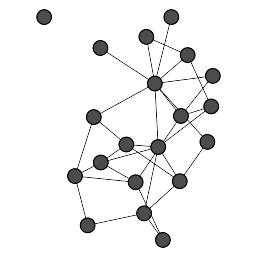}
		\caption{\emph{V} = 20, \emph{E} = 32, \emph{d} = 0.4}
	\end{subfigure}\hspace{1.5cm}
	\begin{subfigure}{0.18\linewidth}
	    \includegraphics[width=\linewidth]{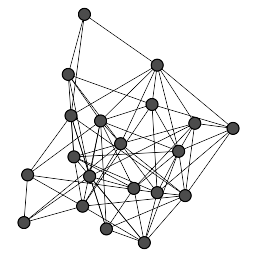}
		\caption{\emph{V} = 20, \emph{E} = 72, \emph{d} = 0.6}
    \end{subfigure} \\
    \begin{subfigure}{0.18\linewidth}
	    \includegraphics[width=\linewidth]{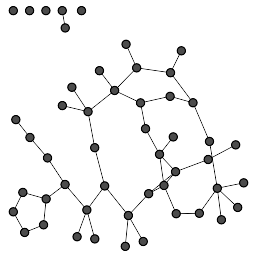}
		\caption{\emph{V} = 50, \emph{E} = 50, \emph{d} = 0.2}
	\end{subfigure}	\hspace{1.5cm}
		 \begin{subfigure}{0.18\linewidth}
	    \includegraphics[width=\linewidth]{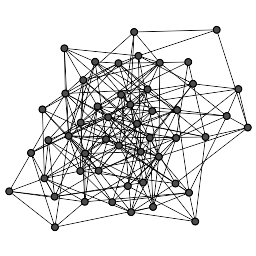}
		\caption{\emph{V} = 50, \emph{E} = 200, \emph{d} = 0.4}
	 \end{subfigure}\hspace{1.5cm}
	 	 \begin{subfigure}{0.18\linewidth}
	    \includegraphics[width=\linewidth]{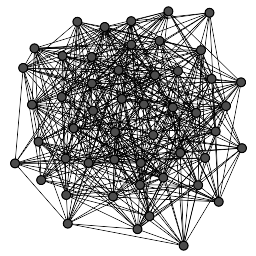}
		\caption{\emph{V} = 50, \emph{E} = 450, \emph{d} = 0.6}
     \end{subfigure} \\
    
	\begin{subfigure}{0.18\linewidth}
	    \includegraphics[width=\linewidth]{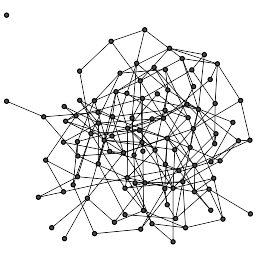}
		\caption{\emph{V} = 100, \emph{E} = 200, \emph{d} = 0.2}
	\end{subfigure}\hspace{1.5cm}
		 \begin{subfigure}{0.18\linewidth}
	    \includegraphics[width=\linewidth]{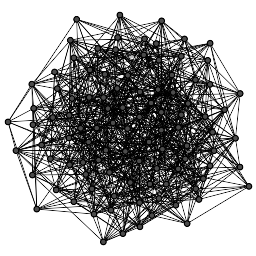}
		\caption{\emph{V} = 100, \emph{E} = 800, \emph{d} = 0.4}
	\end{subfigure}\hspace{1.5cm}
	\begin{subfigure}{0.18\linewidth}
	    \includegraphics[width=\linewidth]{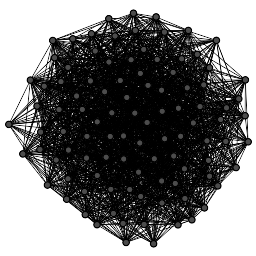}
		\caption{\emph{V} = 100, \emph{E} = 1800, \emph{d} = 0.6}
	\end{subfigure} \\
		\begin{subfigure}{0.18\linewidth}
	    \includegraphics[width=\linewidth]{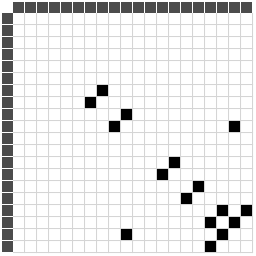}
		\caption{\emph{V} = 20, \emph{E} = 8, \emph{d} = 0.2}
	\end{subfigure}\hspace{1.5cm}
		\begin{subfigure}{0.18\linewidth}
	    \includegraphics[width=\linewidth]{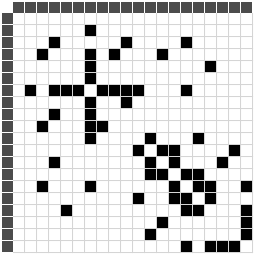}
		\caption{\emph{V} = 20, \emph{E} = 32, \emph{d} = 0.4}
	\end{subfigure}\hspace{1.5cm}
    	 \begin{subfigure}{0.18\linewidth}
	    \includegraphics[width=\linewidth]{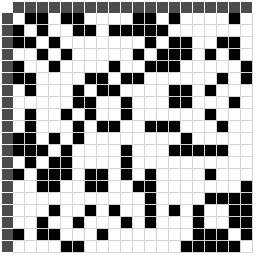}
		\caption{\emph{V} = 20, \emph{E} = 72, \emph{d} = 0.6}
    \end{subfigure} \\
	    \begin{subfigure}{0.18\linewidth}
	    \includegraphics[width=\linewidth]{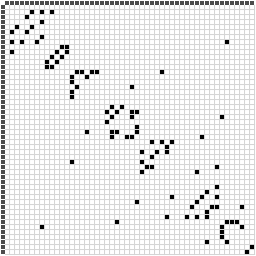}
		\caption{\emph{V} = 50, \emph{E} = 50, \emph{d} = 0.2}
	\end{subfigure}\hspace{1.5cm}
	 	 \begin{subfigure}{0.18\linewidth}
	    \includegraphics[width=\linewidth]{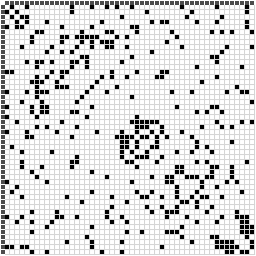}
		\caption{\emph{V} = 50, \emph{E} = 200, \emph{d} = 0.4}
	 \end{subfigure}\hspace{1.5cm}
    	 \begin{subfigure}{0.18\linewidth}
	    \includegraphics[width=\linewidth]{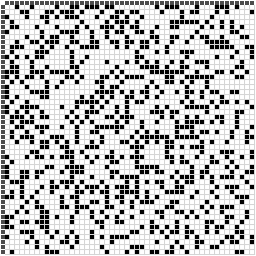}
		\caption{\emph{V} = 50, \emph{E} = 450, \emph{d} = 0.6}
    \end{subfigure} \\
	\begin{subfigure}{0.18\linewidth}
	    \includegraphics[width=\linewidth]{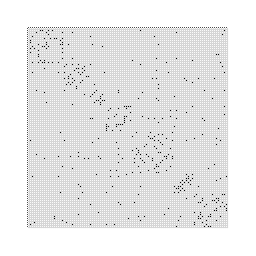}
		\caption{\emph{V} = 100, \emph{E} = 200, \emph{d} = 0.2}
	\end{subfigure}\hspace{1.5cm}
		 \begin{subfigure}{0.18\linewidth}
	    \includegraphics[width=\linewidth]{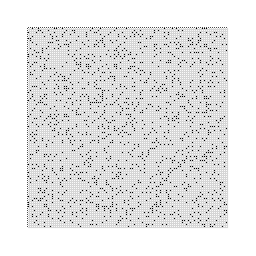}
		\caption{\emph{V} = 100, \emph{E} = 800, \emph{d} = 0.4}
	\end{subfigure}\hspace{1.5cm}
		 \begin{subfigure}{0.18\linewidth}
	    \includegraphics[width=\linewidth]{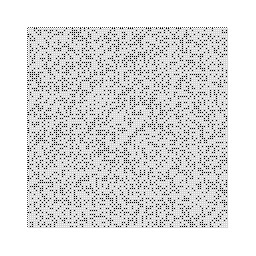}
		\caption{\emph{V} = 100, \emph{E} = 1800, \emph{d} = 0.6}
	\end{subfigure}
		
	\caption{\label{fig_image_example} Examples of uncorrelated graph images for \NL and \MD representations. \emph{V} is the number of nodes, \emph{E} is the number of edges and \emph{d} is the density. \NL layout algorithm is \emph{GEM} and \MD ordering is computed using \emph{Louvain} clustering algorithm.}
\end{figure*} 

To feed the models, we computed a $256 \times 256$ grayscale image for each combination of graph, task and visualization technique (\NL or \MD). 

In \NL diagram, nodes were represented by a circular shape, and each graph has been laid out with the \textit{GEM} force-directed algorithm~\cite{Frick1994AFA}. To make easier the identification of nodes and edges, node size has been set to half the minimum distance between any two nodes in the resulting drawing. 
Concerning \MD diagram, we computed the ordering of nodes by using the \textit{Louvain} clustering algorithm of~\cite{blondel2008fast}. This emphasizes the community structures of the graph even if assessing such a topological information is not necessary (given our tasks and the random data generation model). In this representation, nodes were represented by squares. For readability purposes, we drew the grid of the matrix with lines of width \num{1} pixel. Finally, row and column size (resp. height and width) has been set as large as possible to maximize space usage.

Nodes appear larger in \NL than in \MD (see Figure~\ref{fig_image_example}) as we maximized the space usage in the computed images and due to the different properties of the two visualization techniques.

\paragraph{\textbf{Deep convolutional network architectures}}
\label{ghoniemDeepConv}
\label{sec_models}
For this experiment, we have selected two common convolutional network architectures. On one hand, we selected a simple yet efficient architecture for solving \textit{simple} problems. That architecture, called \lenet~\cite{lecun1998gradient}, contains \num{8} layers and has been initially designed to recognize hand-written digits from small $28 \times 28$ images. 
On the other hand, we selected a much more complex one, called \linebreak\vgg~\cite{simonyan2014very}, that has been shown to perform well with \textit{complex} image classification problems.~\cite{haehn2018evaluating} mentioned that VGG architecture provides the best results out of their four tested architectures.
It seems therefore to be a good candidate to run comparison between visualization techniques.   
That architecture contains \num{21} layers and has been initially designed for large-scale image classification problems.

We have chosen a classification approach over a regression one (\emph{i.e.}, models predict a class instead of a value) to generalize them for other tasks in the future. 
For the same reason, the output layer has also been oversized (\emph{i.e.} some classes do not appear in the data) compared to the number of classes to predict to be more dataset independent.

We trained a model for each combination of architecture-task-visualization technique.
Each network has been completely trained (as opposed to pre-trained networks) and 
no weight was shared between the networks of a same architecture for the different tasks (as it could be done by~\cite{haleem2018evaluating}).
Each model has been trained during \num{100} epochs using negative log likelihood loss and the Adadelta optimizer from~\cite{zeiler2012adadelta} that has the advantage of requiring no  manual tuning of the learning rate.

 \begin{table}[!tb]
	
	\centering
	\caption{\label{tab_learning}$R^2$ scores of the best models on \accuracy metric (see Section~\ref{sec:eval_metrics}) for each combination of task, network architecture, and visualization technique ($1.0$ values are due to rounding).
	}
	\resizebox{0.85\linewidth}{!}{
		
		\begin{tabular}{lp{2cm}rr}
			\toprule
		    
			\textbf{Task} & \textbf{Network} &  \multicolumn{2}{c}{\textbf{$R^2$ score per vis. technique}} \\
			& \textbf{Architecture}&  \textbf{\MD}    &    \textbf{\NL} \\
			\midrule
			
			\multirow{2}{*}{\Task{0}} & \lenet &  \num{1.0} &  \num{0.73}  \\
									  & \vgg & \num{1.0}  &   \num{0.97} \\
			
			\midrule
			\multirow{2}{*}{\Task{1}} & \lenet &  \num{1.0} &  \num{0.99}  \\
									& \vgg & \num{1.0} &  \num{1.0} \\
			
			\midrule
			\multirow{2}{*}{\Task{2}} & \lenet &  \num{0.98} &  \num{0.97}  \\
									& \vgg & \num{0.98} &   \num{0.97} \\
			\midrule
			\multirow{2}{*}{\Task{3}} & \lenet &  \num{-1.48} &  \num{0.46}  \\
							        & \vgg & \num{-1.18e-6} &   \num{-1.18e-6} \\
							
			\bottomrule
		\end{tabular}
	}
\end{table}

\begin{table}[!htb]
\caption{\label{tab_summary} Summary of the results obtained for each combination of tasks, network and evaluation metric. 
For each triplet, the best epoch is obtained from the validation database while the performance is obtained from the test database.
The best configuration over the rows is presented in bold.
}

\centering

\resizebox{\linewidth}{!}{

	\begin{tabular}{lll
		S[detect-weight,group-minimum-digits = 4,table-format=5.0]
		S[detect-weight,group-minimum-digits = 4,table-format=5.0]
		rr}
		\toprule
				 &            & {} & \multicolumn{2}{c}{\textbf{Performance}} & \multicolumn{2}{c}{\textbf{Best epoch}} \\
				 &            & \textbf{Representation} &           \textbf{\MD} &            \textbf{\NL} &   \textbf{\MD} & \textbf{\NL} \\
		\textbf{Task} & \textbf{Evaluation} & \textbf{Network} &               &                &       &     \\
		\midrule
	
		\multirow{8}{*}{{\Task{0}}} & \multirow{2}{*}{{\accuracy}} & \lenet &     \best{1.0} &      \num{0.1} &   \best  11 &  33 \\
		&            & \vgg &     \best{1.0} &     \num{0.15} &  \best  18 &  25 \\
\cline{2-7}
		& \multirow{2}{*}{{\deltametric}} & \lenet &     \best{1.0} &     \num{0.86} &    \best 5 &  47 \\
		&            & \vgg &     \best{1.0} &     \num{0.97} &  \best  13 &  24 \\
\cline{2-7}
		& \multirow{2}{*}{{\mse}} & \lenet &     \best{0.0} &    \num{143.6} &   \best  11 &  47 \\
		&            & \vgg &     \best{0.0} &    \num{19.23} &   \best 18 &  25 \\
\cline{2-7}
		& \multirow{2}{*}{{\percent}} & \lenet &     \best{1.0} &     \num{0.71} &   \best  5 &  47 \\
		&            & \vgg &     \best{1.0} &     \num{0.82} &  \best  14 &  24 \\

\midrule
\midrule

\multirow{8}{*}{{\Task{1}}} & \multirow{2}{*}{{\accuracy}} & \lenet &     \best{1.0} &     \num{0.14} &  \best  17 &  50 \\
		&            & \vgg &    \best{0.95} &     \num{0.19} &    69 & \best 49 \\
\cline{2-7}
		& \multirow{2}{*}{{\deltametric}} & \lenet &     \best{1.0} &      \num{0.4} & \best   17 &  76 \\
		&            & \vgg &    \best{0.96} &     \num{0.51} &    86 & \best 49 \\
\cline{2-7}
		& \multirow{2}{*}{{\mse}} & \lenet &    \best{0.04} &  \num{1327.56} & \best   17 &  38 \\
		&            & \vgg &  \best{174.04} &   \num{606.15} &    59 & \best  55 \\
\cline{2-7}
		& \multirow{2}{*}{{\percent}} & \lenet &     \best{1.0} &     \num{0.61} &   \best 17 &  78 \\
		&            & \vgg &    \best{0.95} &     \num{0.74} &    86 & \best  49 \\

\hline
\hline

\multirow{8}{*}{{\Task{2}}} & \multirow{2}{*}{{\accuracy}} & \lenet &    \best{0.29} &     \num{0.25} &    16 & \best  15 \\
		&            & \vgg &    \best{0.28} &     \num{0.26} &  \best  19 &  36 \\
\cline{2-7}
		& \multirow{2}{*}{{\deltametric}} & \lenet &     \best{1.0} &      \best{1.0} &     7 & \best  5 \\
		&            & \vgg &     \best{1.0} &      \best{1.0} &    19 & \best   6 \\
\cline{2-7}
		& \multirow{2}{*}{{\mse}} & \lenet &    \best{2.64} &     \num{4.92} & \best   10 &  64 \\
		&            & \vgg &    \best{2.64} &     \num{3.37} &  \best   19 &  31 \\
\cline{2-7}
		& \multirow{2}{*}{{\percent}} & \lenet &    \best{0.66} &     \num{0.56} &    \best 10 &  24 \\
		&            & \vgg &    \best{0.64} &     \num{0.63} &  \best  19 &  31 \\

\hline
\hline

\multirow{2}{*}{{\Task{3}}} & \multirow{2}{*}{{\accuracy}} & \lenet &    \num{0.34} &     \best{0.79} &    \best{1} & \num{76} \\
                                                         &   & \vgg &    \num{0.32} &     \num{0.32} &    \best{1} & \num{3} \\
\cline{2-7}
		& \multirow{2}{*}{{\mse}} & \lenet &    \num{1.68} &  \best{0.27} & \best{1} &  \num{84} \\
		                    &   & \vgg &    \num{0.68} &     \num{0.68} &    \best{1} & \num{3} \\
		\bottomrule
		\end{tabular}
}
\end{table}

 \begin{table}[!htb]
	\sisetup{round-mode=places}

	\centering
	\caption{\label{tab_md_best_nl} Statistical validation of the hypotheses.
	For each task $\times$ representation, the mean of the absolute difference between the ground-truth and the prediction is shown for \MD and \NL over all the possibilities graph $\times$ representation; the p-value of the Wilcoxon test is also provided (0 values are due to rounding issues).
	}
\resizebox{0.90\linewidth}{!}{

	\begin{tabular}{lp{2cm}rrr}
		\toprule
		  &           &    \multicolumn{2}{c}{\textbf{Mean of abs. diff.}}      &       \textbf{p-value} \\
		\textbf{Task} & \textbf{Model selection criterion} &  \textbf{\md}     &     \textbf{\nl}   &               \\
		\midrule

		\multirow{4}{*}{\Task{0}} & {\mse} &  \best 0.00 &   4.64 &  \num{0.000000e+00} \\
		& \accuracy & \best  0.00 &   4.86 &  \num{0.000000e+00} \\
		& \percent & \best 0.05 &   4.66 &  \num{0.000000e+00} \\
		& \deltametric & \best 0.05 &   4.66 &  \num{0.000000e+00} \\
	\cline{1-5}
	\multirow{4}{*}{\Task{1}} & {\mse} & \best 0.98 &  20.15 &  \num{0.000000e+00} \\
		& \accuracy & \best 1.00 &  20.38 &  \num{0.000000e+00} \\
		& \percent & \best 1.01 &  19.90 &  \num{0.000000e+00} \\
		& \deltametric & \best 1.01 &  20.15 &  \num{0.000000e+00} \\
	\cline{1-5}
	\multirow{4}{*}{\Task{2}} & {\mse} & \best 1.16 &   1.46 &  \num{3.034237e-49} \\
		& \accuracy & \best 1.17 &   1.53 &  \num{9.854858e-60} \\
		& \percent & \best 1.16 &   1.44 &  \num{8.759492e-39} \\
		& \deltametric & \best 1.24 &   1.66 &  \num{8.993975e-78} \\
	\cline{1-5}
    \multirow{2}{*}{\Task{3}} & {\mse} & \num{0.84} &   \best{0.46} &  \num{6.45e-239} \\
		& \accuracy & \num{0.84} &  \best{0.47} &  \num{9.07e-203} \\

    	\bottomrule
		\end{tabular}
}
 \end{table}

\begin{figure*}[!tb]
 \centering

	\begin{tabular}{|c|c|c|c|c|c|}
		\hline
		& & \task{0} & \task{1} & \task{2} & \task{3} \\
		\hline
		\multirow{2}{*}{\begin{sideways}\lenet\end{sideways}} 
		&
	\begin{sideways}{\hspace{1cm}\MD}\end{sideways} & 

		{\includegraphics[width=.17\linewidth]{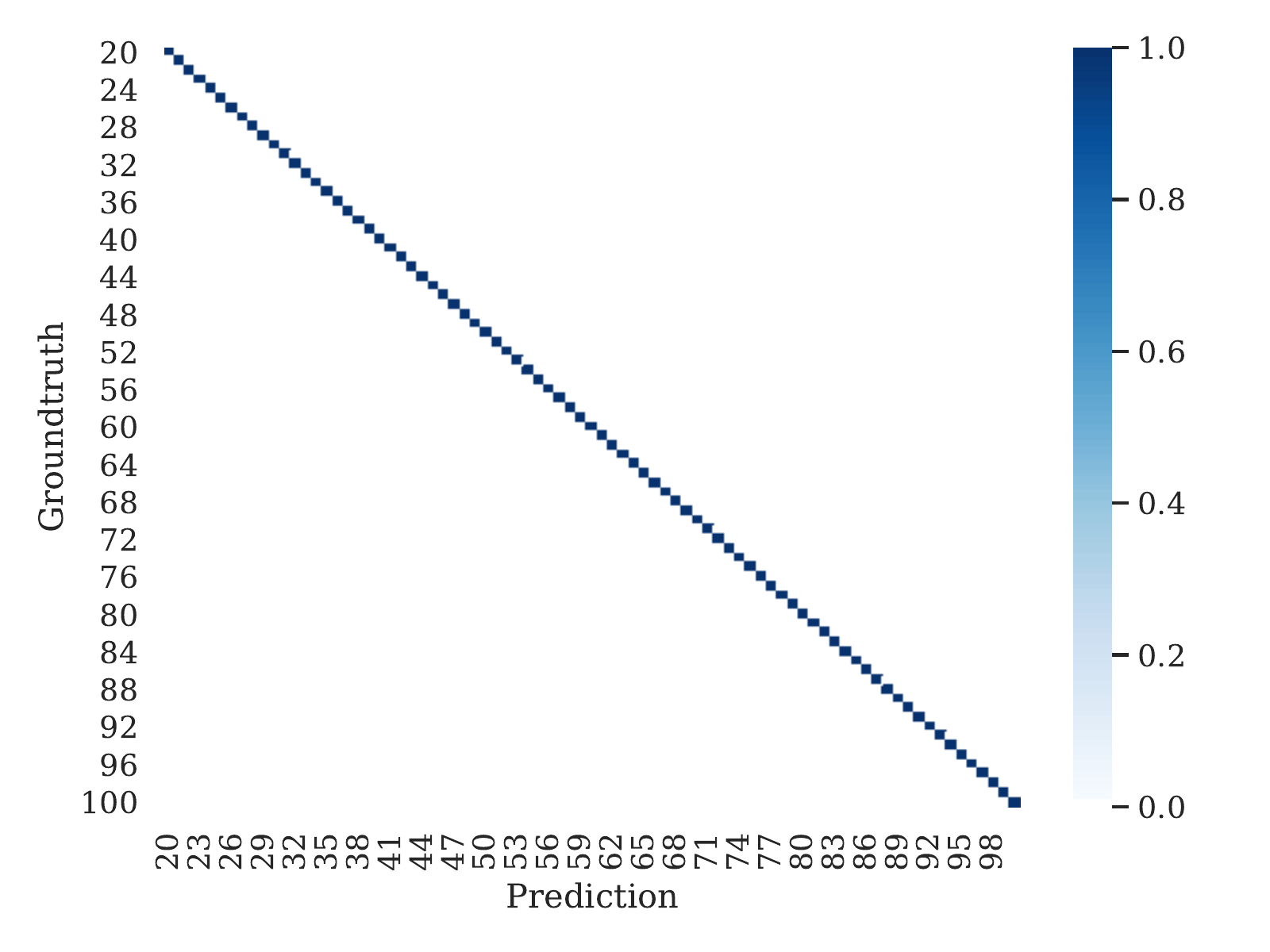}}&%
		{\includegraphics[width=.17\linewidth]{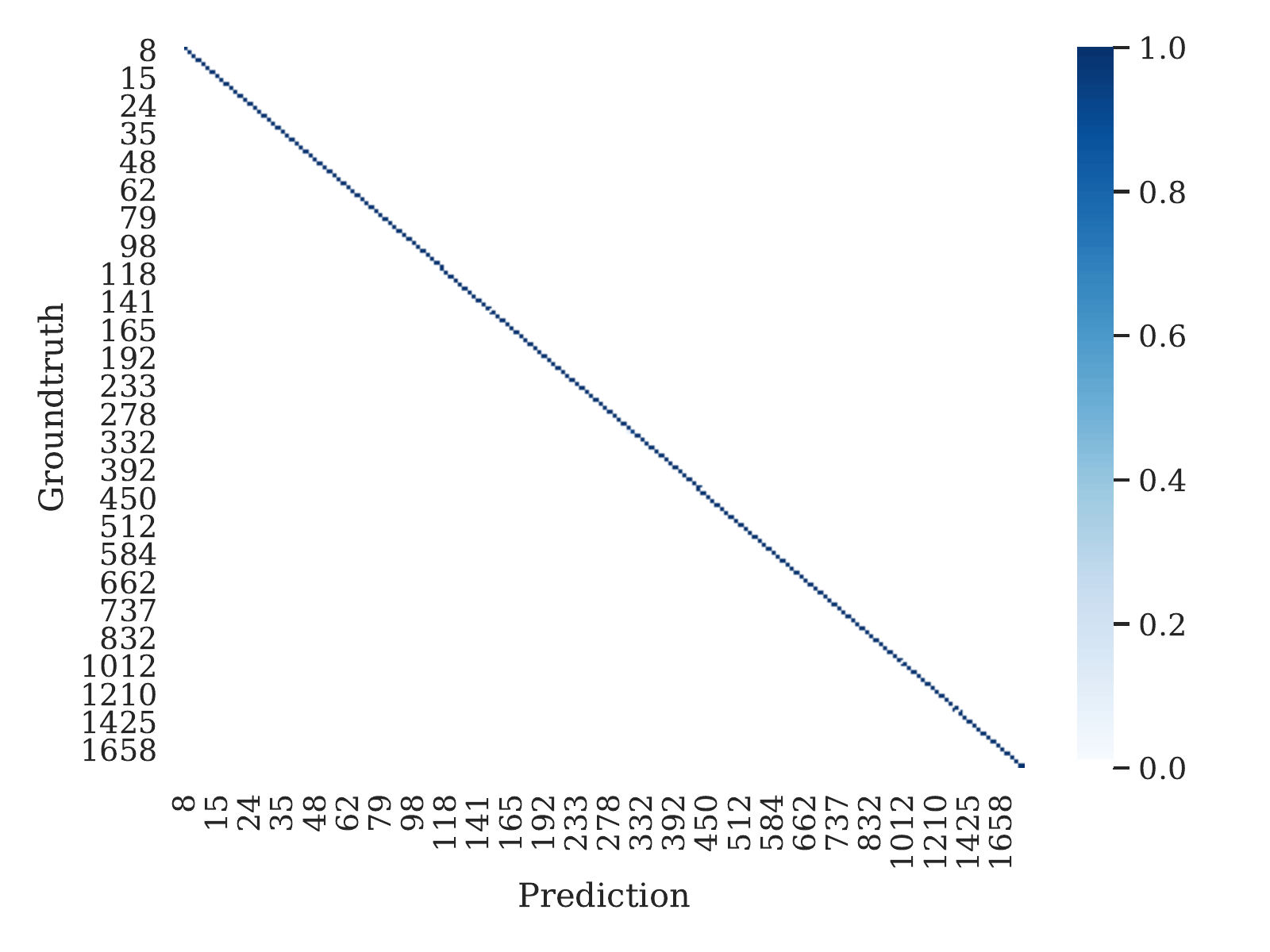}}&%
		{\includegraphics[width=.17\linewidth]{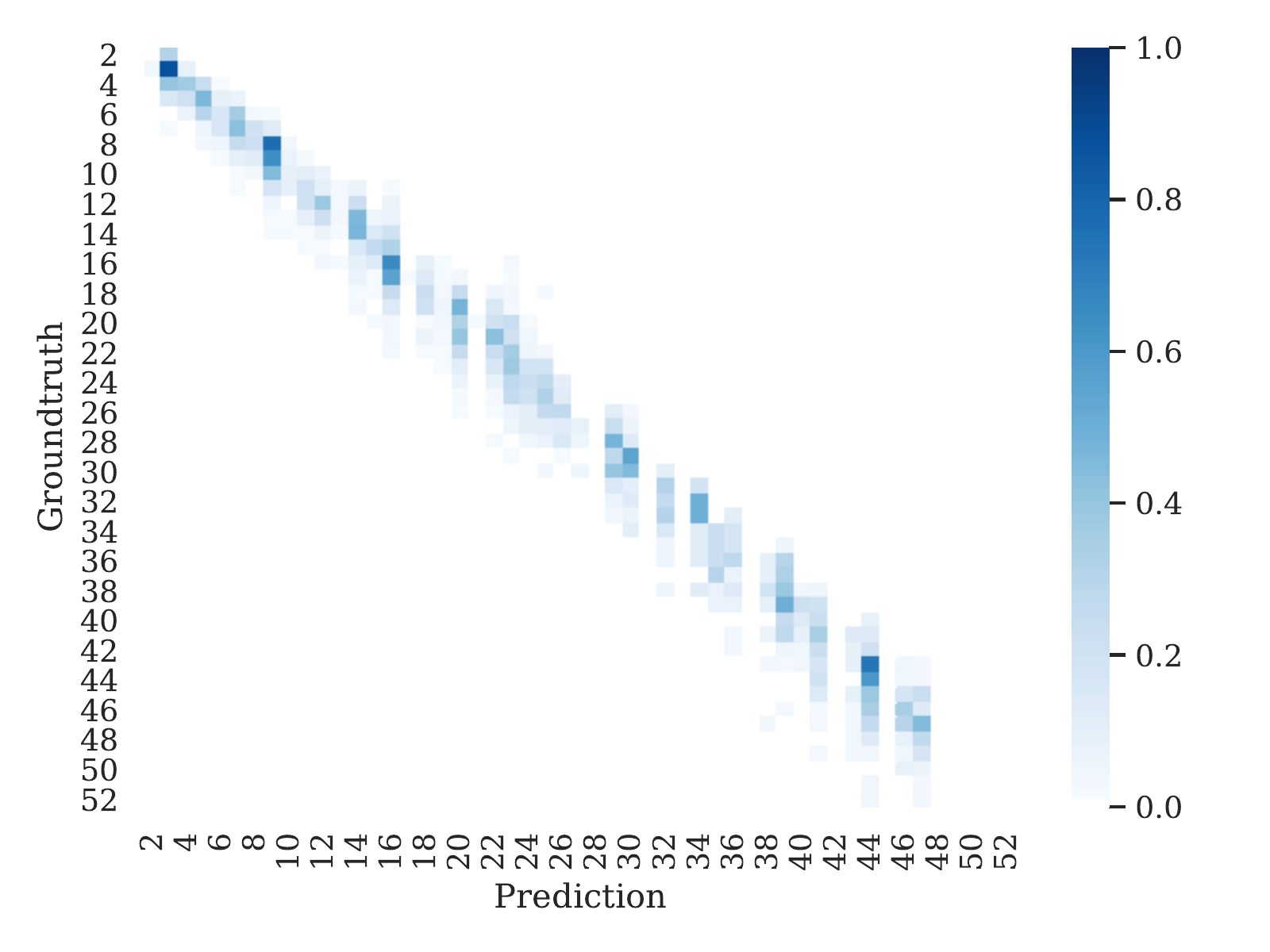}}&%
		{\includegraphics[width=.17\linewidth]{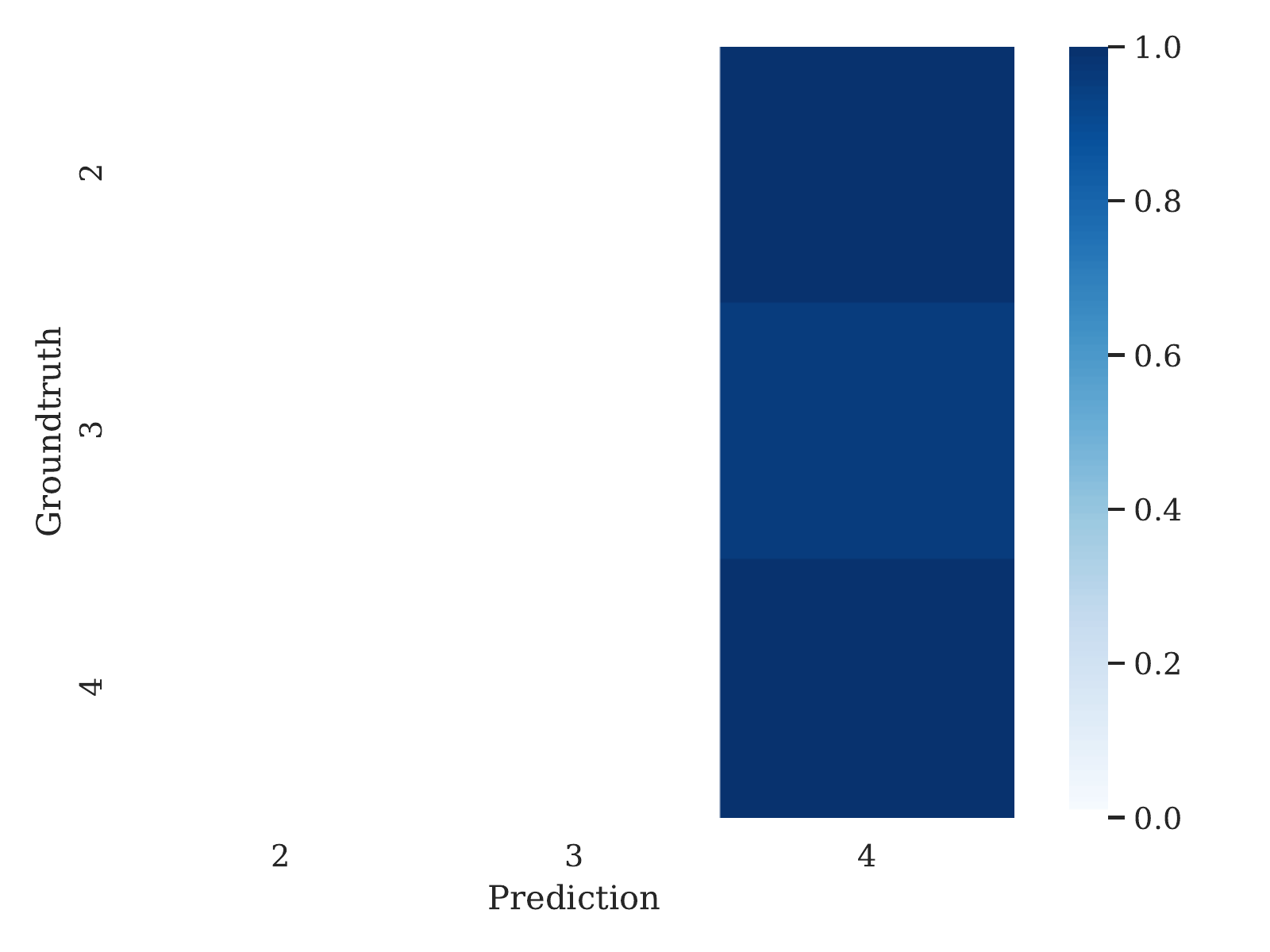}}\\%

		\cline{2-6}
		&
	\begin{sideways}{\hspace{1cm}\NL}\end{sideways} & 
		{\includegraphics[width=.17\linewidth]{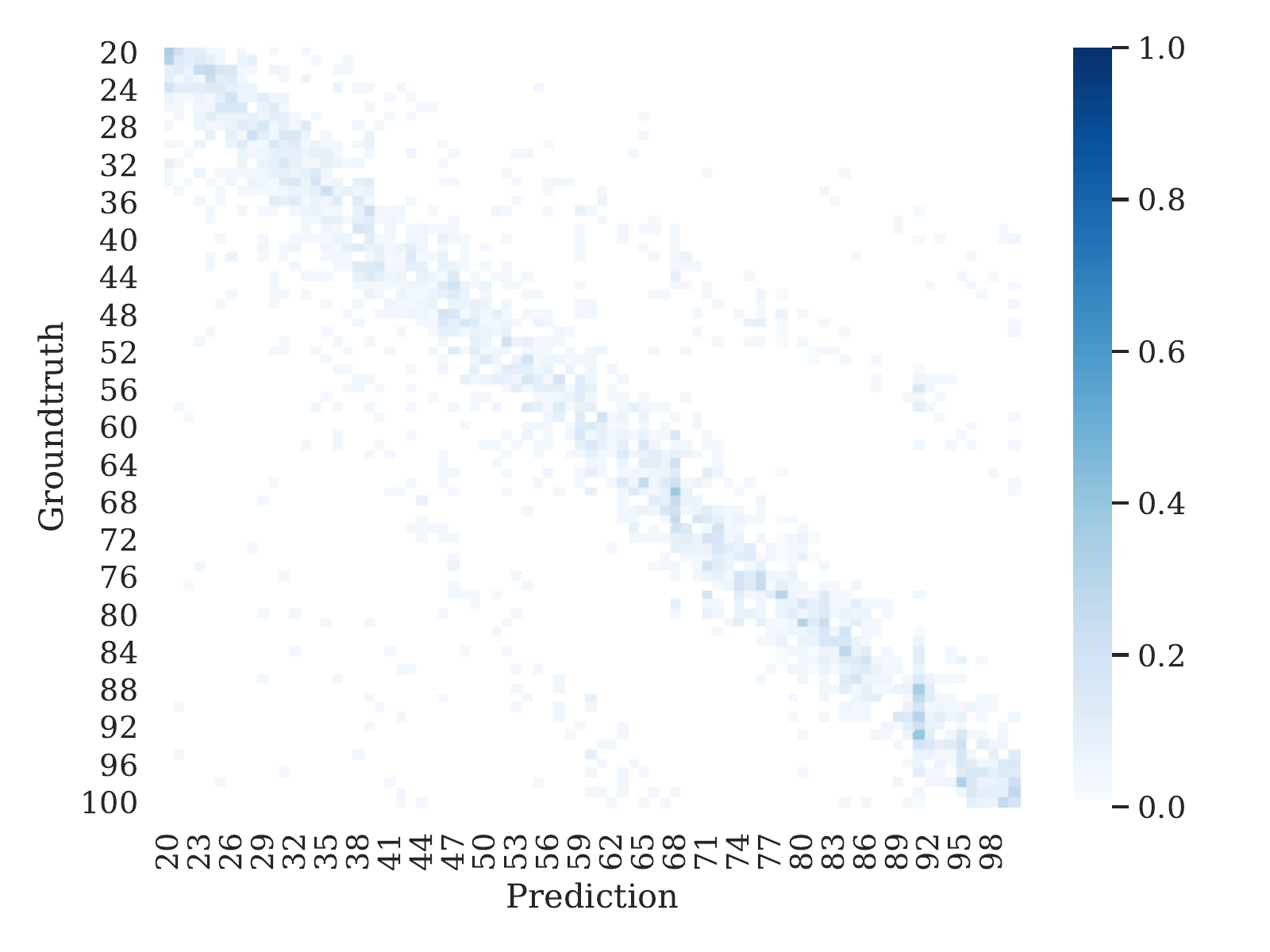}}&%
		{\includegraphics[width=.17\linewidth]{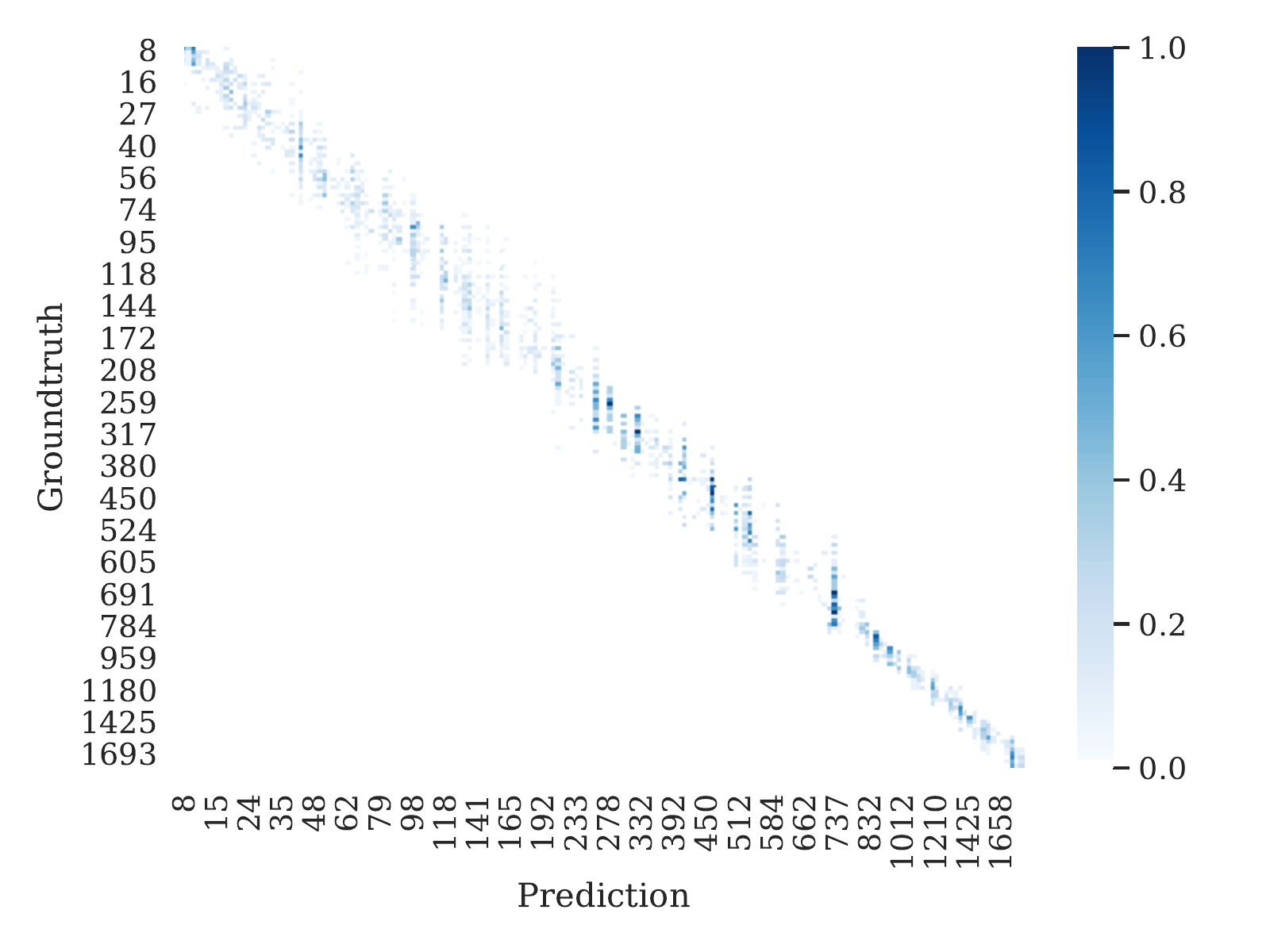}}&%
		{\includegraphics[width=.17\linewidth]{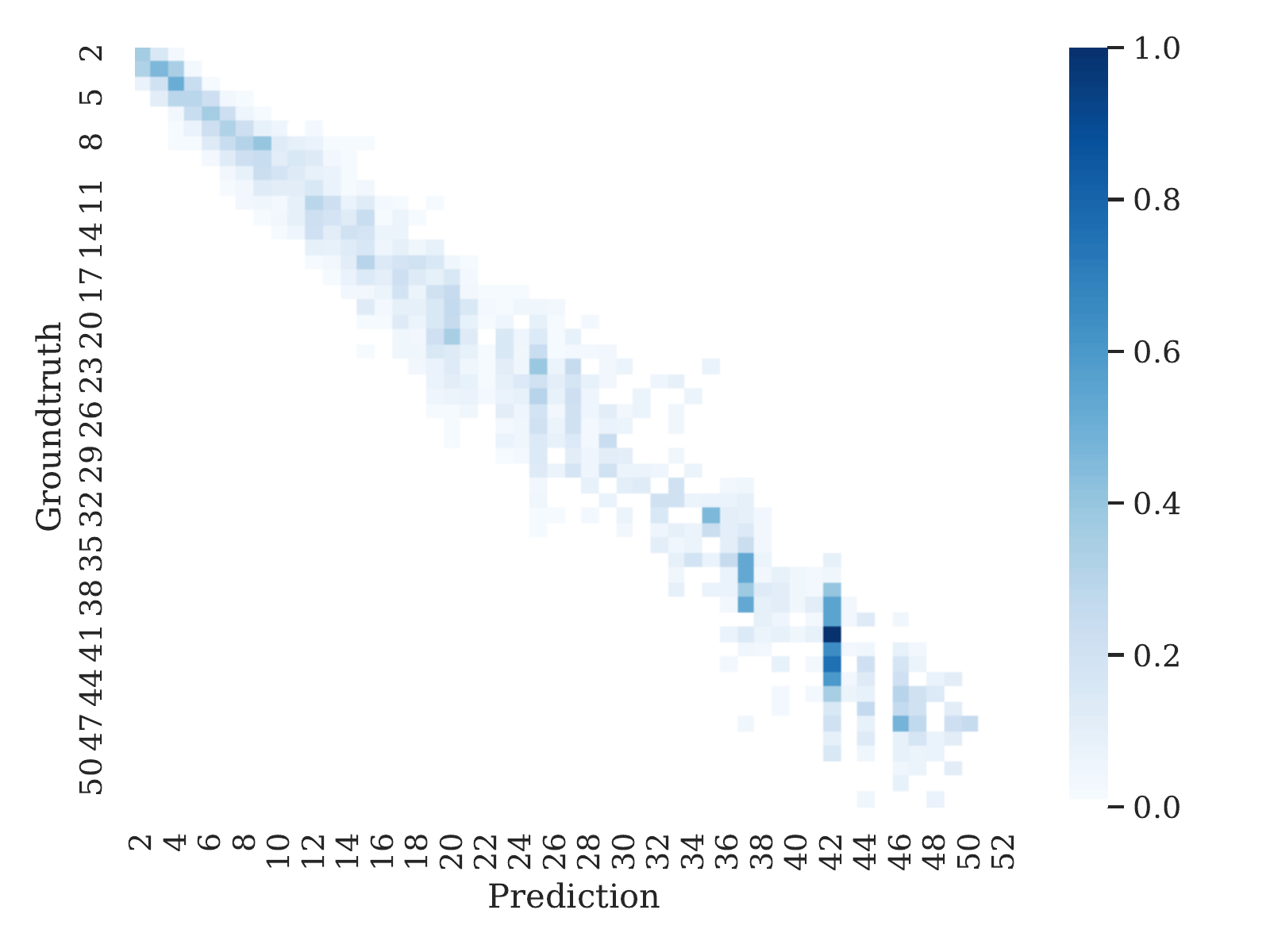}}&%
		{\includegraphics[width=.17\linewidth]{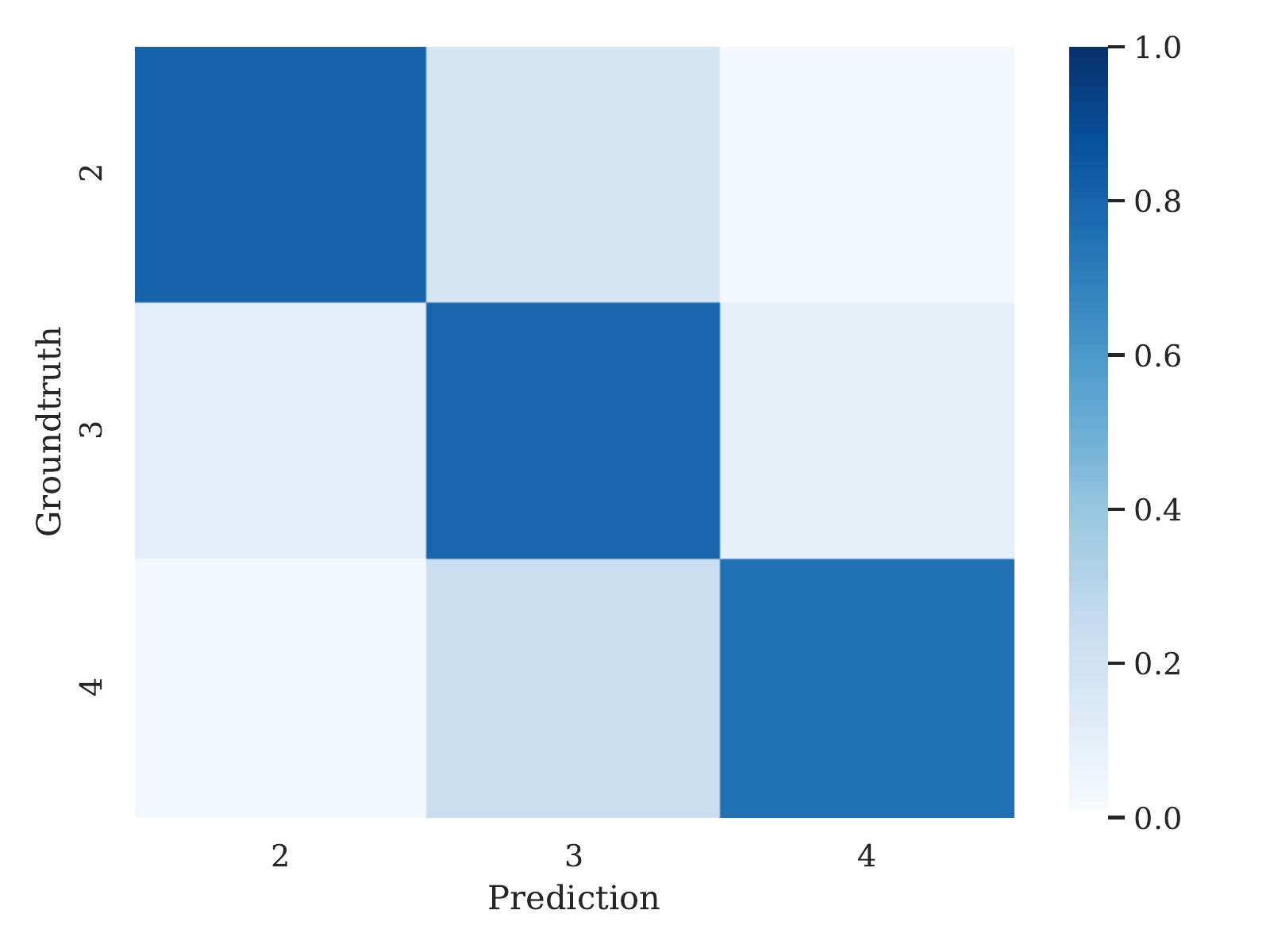}}\\%

	\hline
\multirow{2}{*}{\begin{sideways}\vgg\end{sideways}} 
&
\begin{sideways}{\hspace{1cm}\MD}\end{sideways} & 

{\includegraphics[width=.17\linewidth]{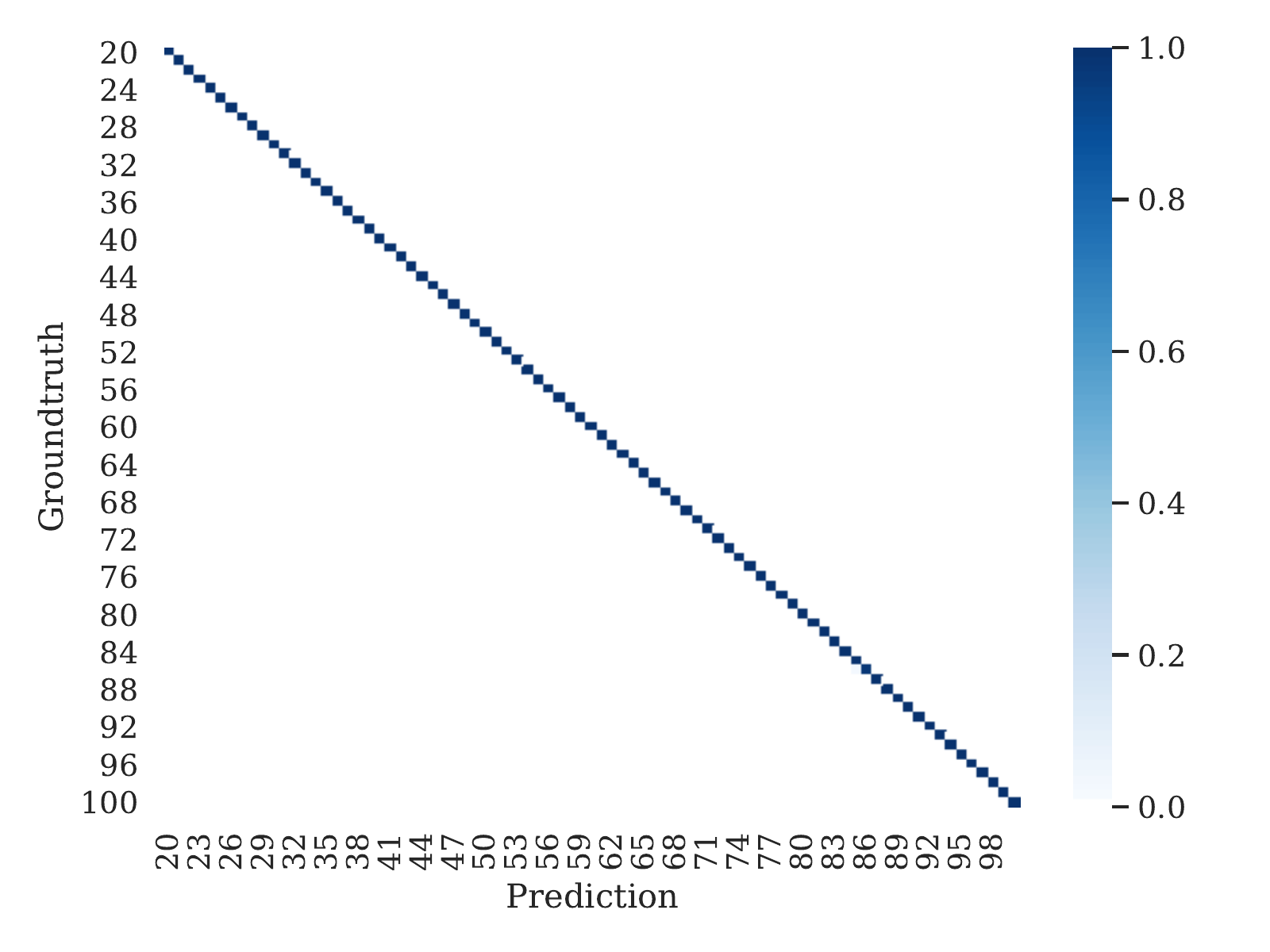}}&%
{\includegraphics[width=.17\linewidth]{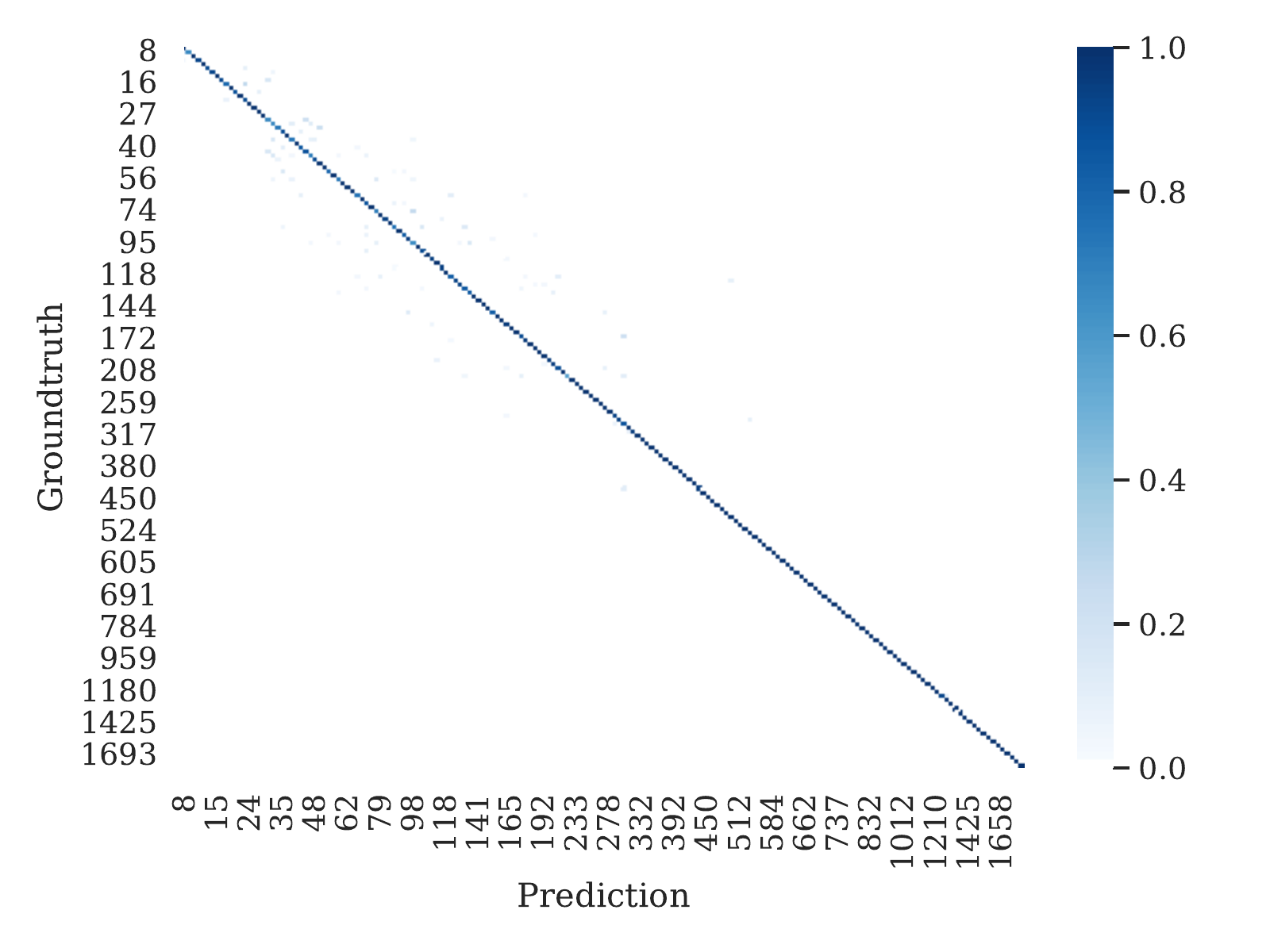}}&%
{\includegraphics[width=.17\linewidth]{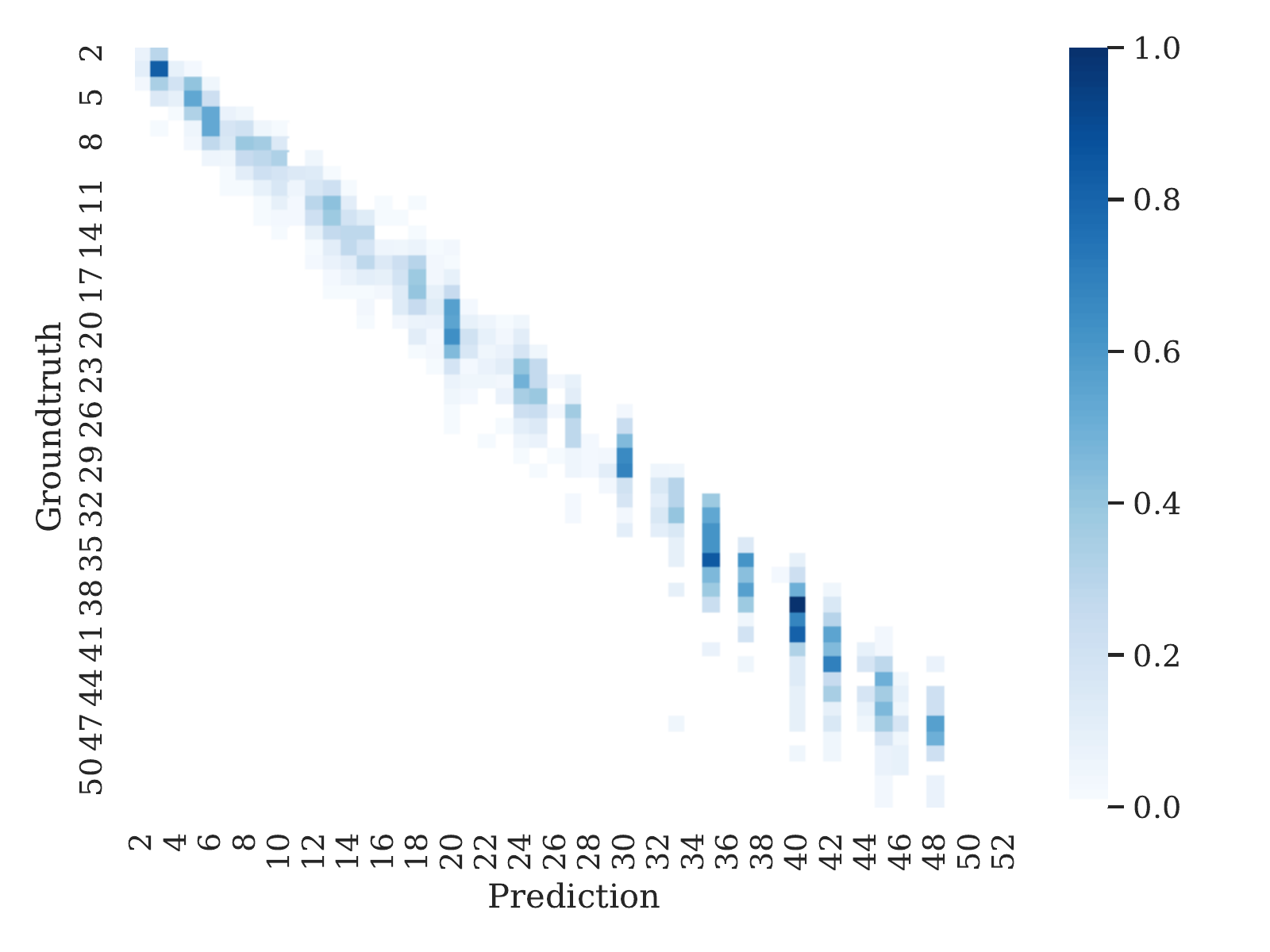}}&%
{\includegraphics[width=.17\linewidth]{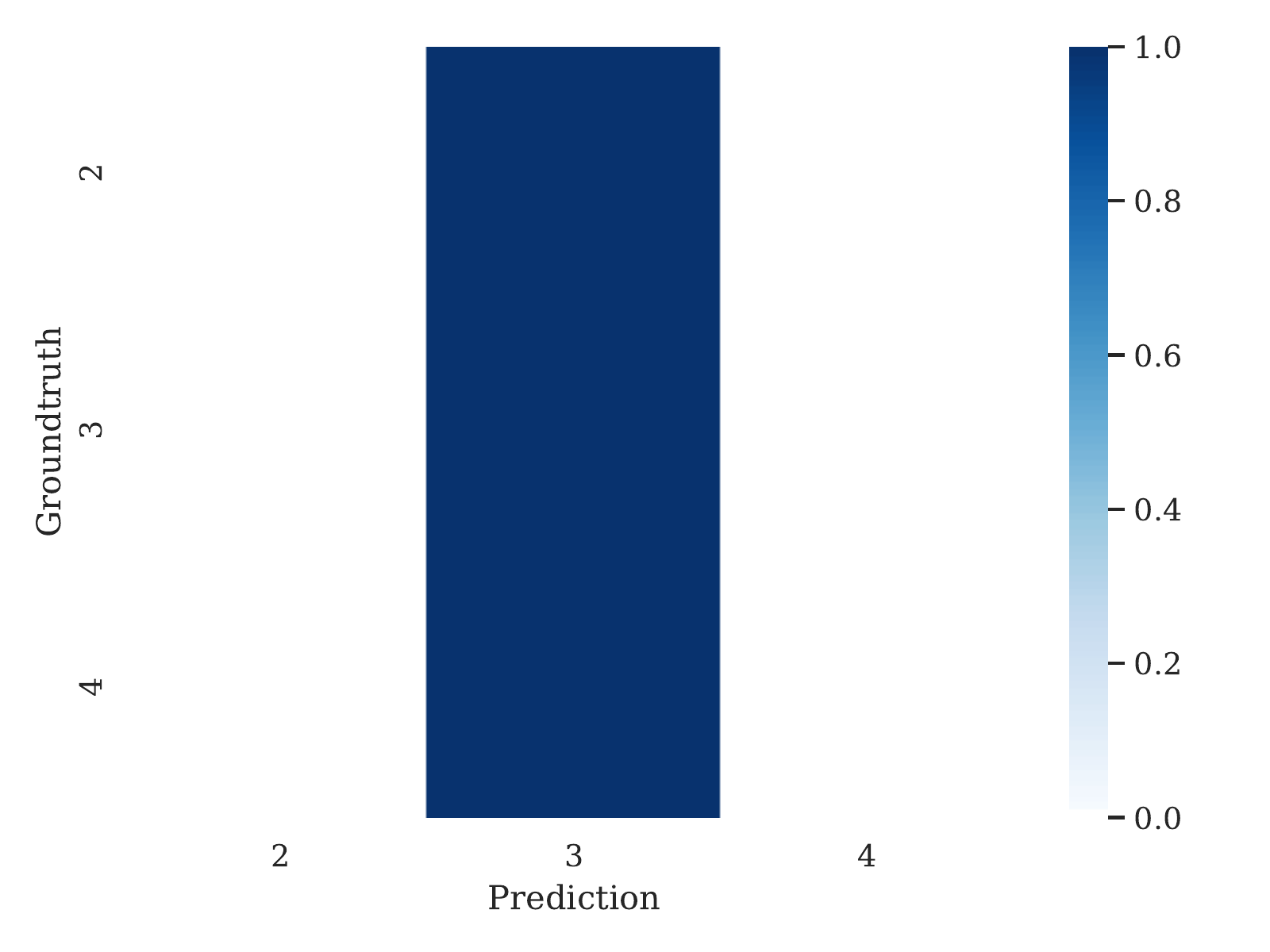}}\\%

\cline{2-6}
&
\begin{sideways}{\hspace{1cm}\NL}\end{sideways} & 
{\includegraphics[width=.17\linewidth]{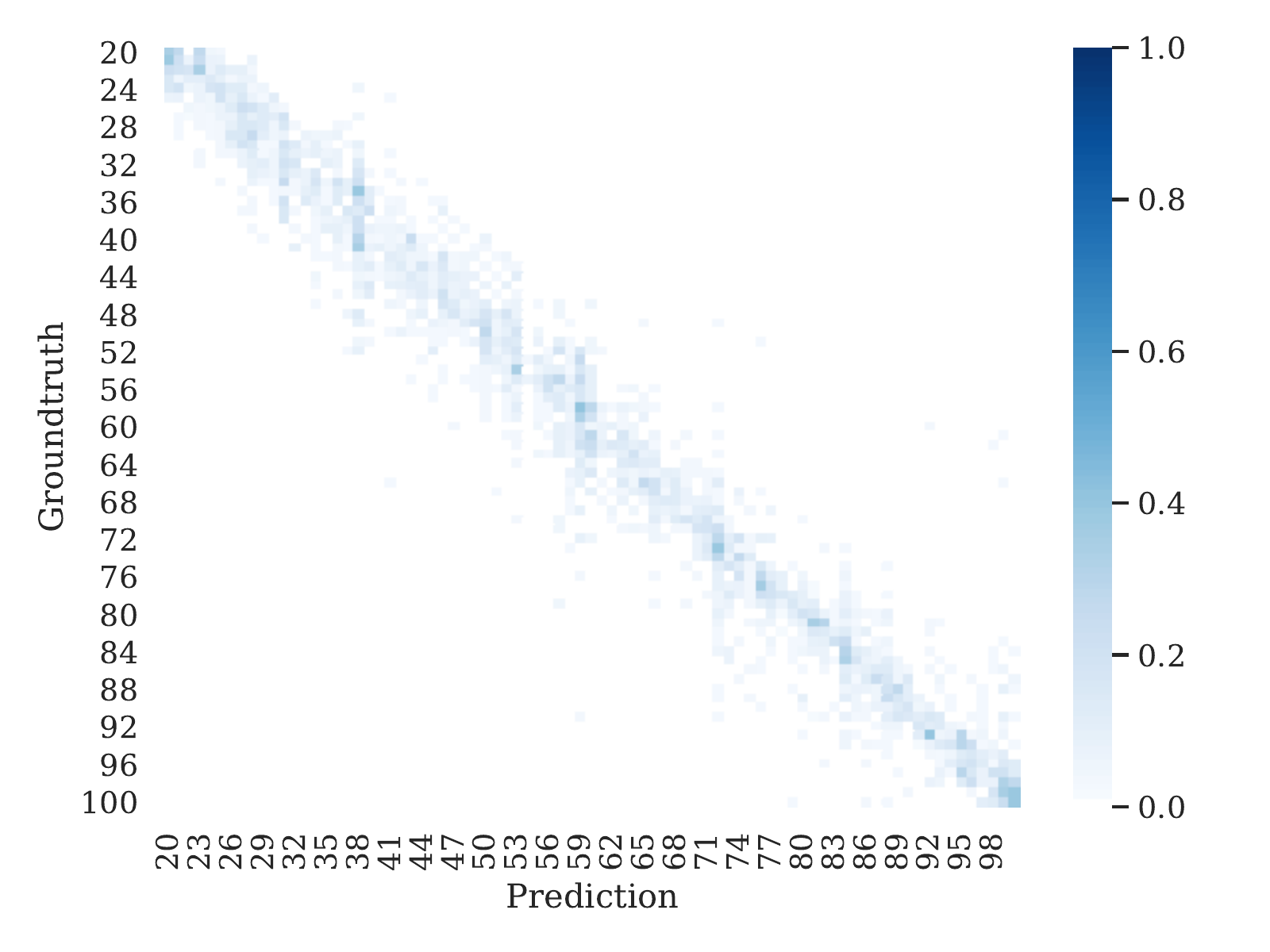}}&%
{\includegraphics[width=.17\linewidth]{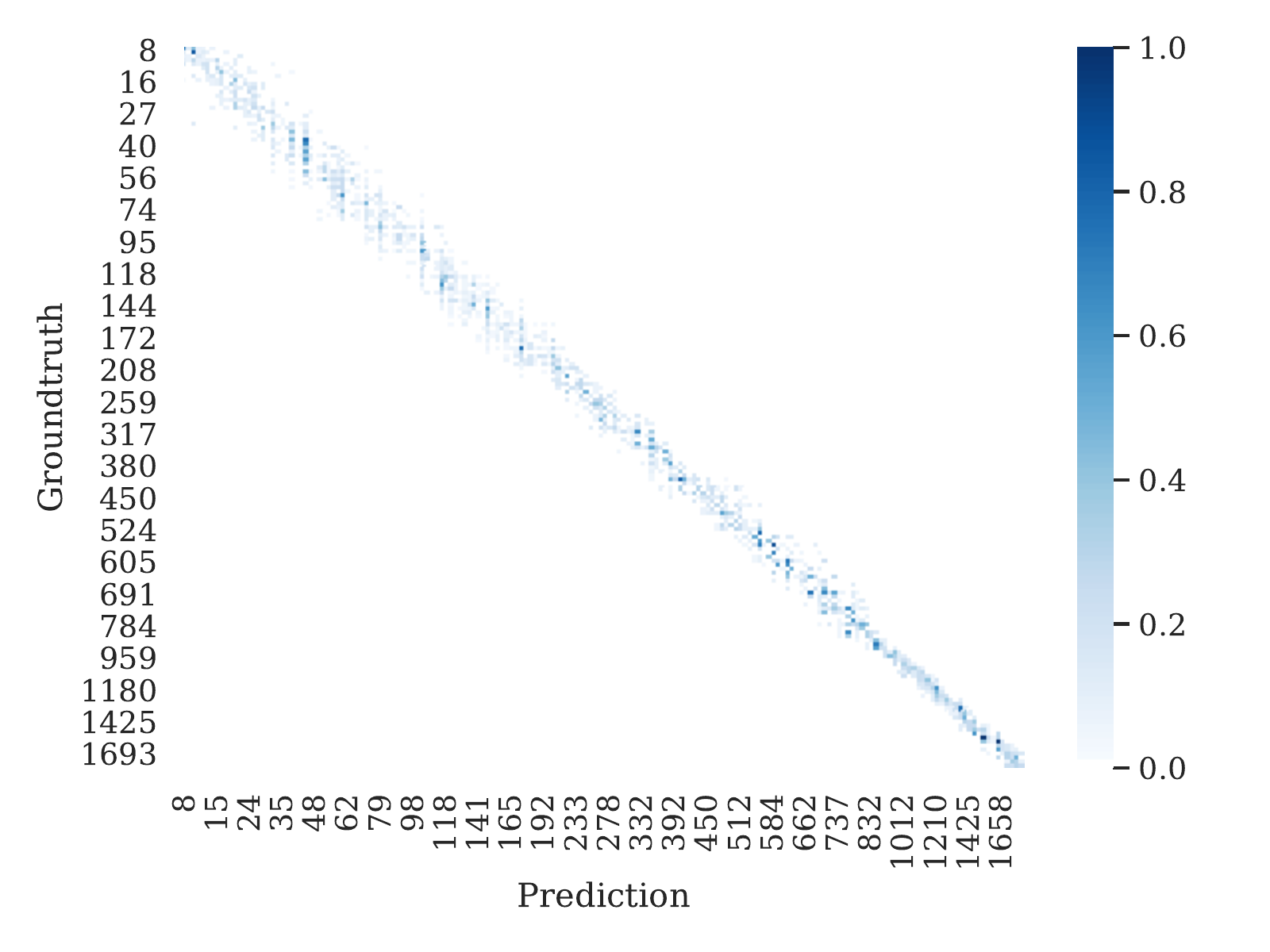}}&%
{\includegraphics[width=.17\linewidth]{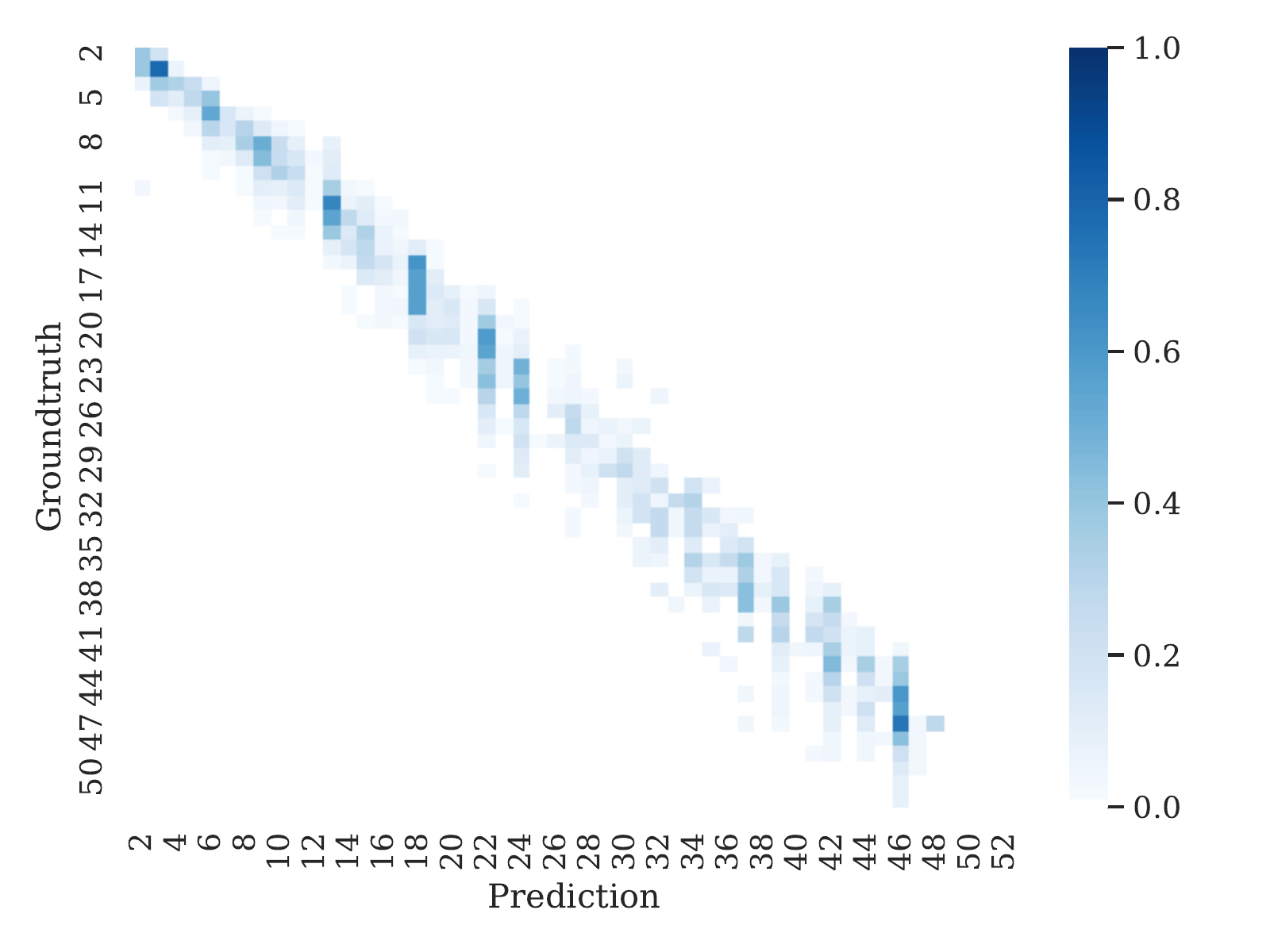}}&%
{\includegraphics[width=.17\linewidth]{images/kobourov/confusion_Task0_vgg16_NL_mse.pdf}}\\%

		\hline
	\end{tabular}
  \caption{Confusion matrices with the best model on \mse metric for each task and representation, computed on the test database. \lenet and \vgg models have been trained to solve the three considered graph visualization tasks on both \MD and \NL diagrams..
  	In a perfect performing system, the diagonal is dark blue while the other cells are white. 
  	\vgg seems to perform better than \lenet on \NL representation while the opposite can be observed for \MD representation.  	
}
  
\label{fig:vgg16_matrix}
\end{figure*}

\subsubsection{Results}
\label{sec_results}

We are interested in verifying if the models could learn, how they perform and how they compare between the two graphs representations to validate our hypothesis. Another interesting point is the learning convergence speed.
This section presents these different aspects.

\paragraph{\textbf{Learning feasibility}}
As mentioned in Section~\ref{sec_methodology}, it is mandatory to determine whether the models could learn how to solve the tasks. Even if we use classification models in this experiment, the models solve a regression problem as the goal is to predict a value (\emph{e.g.} the number of nodes). We therefore used $R^2$ metric to assess how well the models did learn. 
Table~\ref{tab_learning} shows the $R^2$ scores of the best models for each combination of the considered parameters in our experiment. These $R^2$ scores range from \num{0.73} (for the \Task{0} on \nl with \lenet architecture) to \num{1.}. 
It shows that both \lenet and \vgg have learned how to solve the three tasks with each of the visualization techniques.

\paragraph{\textbf{Models performances and comparison}}
\label{GhoniemPerfComp}
Table~\ref{tab_summary} summarizes the classification results using different metrics computed on $V_{test}$ by the models performing the best on $V_{validation}$.

 Our hypothesis was that automatic evaluation should reproduce the results of~\cite{ghoniem2004comparison,ghoniem2005readability}, \textit{i.e.} \MD should outperform \NL for each of the three tasks.
To validate our hypothesis on each task, we selected for each metric (\mse, \accuracy, \percent, \deltametric) the best models (one for each architecture) for each visualization technique. For instance, Figure~\ref{fig:vgg16_matrix} shows the confusion matrix of the best models selected using \mse for each visualization technique. 
Then we evaluated how well the best models approximate the ground-truths by computing for each graph the absolute difference between the ground-truth and their predictions (see Table~\ref{tab_md_best_nl}).
For each selection criterion (\textit{i.e.} metric), we compared these absolute differences for \MD and \NL.
We used the Wilcoxon test that verifies if these two sets of data share the same distribution. 
If the p-value is lower than $0.05$, we can consider that they are different in favor of the one having the smallest mean absolute difference, 
otherwise no statement can be made.
This process is repeated four times by using the best models obtained with the evaluation metrics (\mse, \accuracy, \percent, \deltametric).
Table~\ref{tab_md_best_nl} shows that the p-value is systematically $<<0.05$ (so the distributions are different) for the \md representation;
Kruskal Wallis test provides the same results. 	
These results validate the hypothesis \textit{H1:} \MD outperforms \NL for the three considered tasks.

%
%
%
\begin{figure}[!tb]
	\centering

	 \begin{subfigure}{0.49\linewidth}
	    \includegraphics[width=\linewidth]{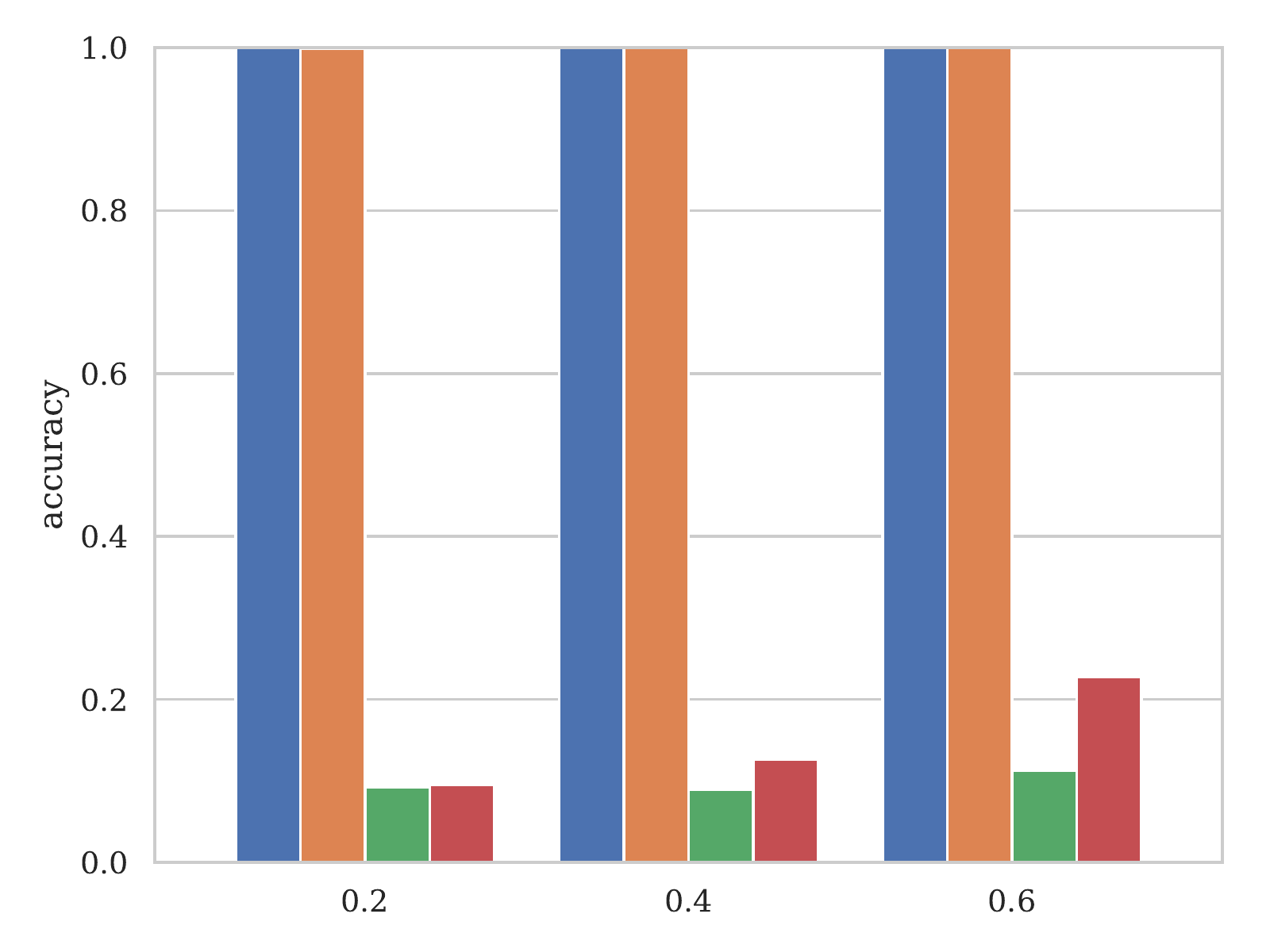}
		\caption{\task{0}.}
	 \end{subfigure}
	 \begin{subfigure}{0.49\linewidth}
	    \includegraphics[width=\linewidth]{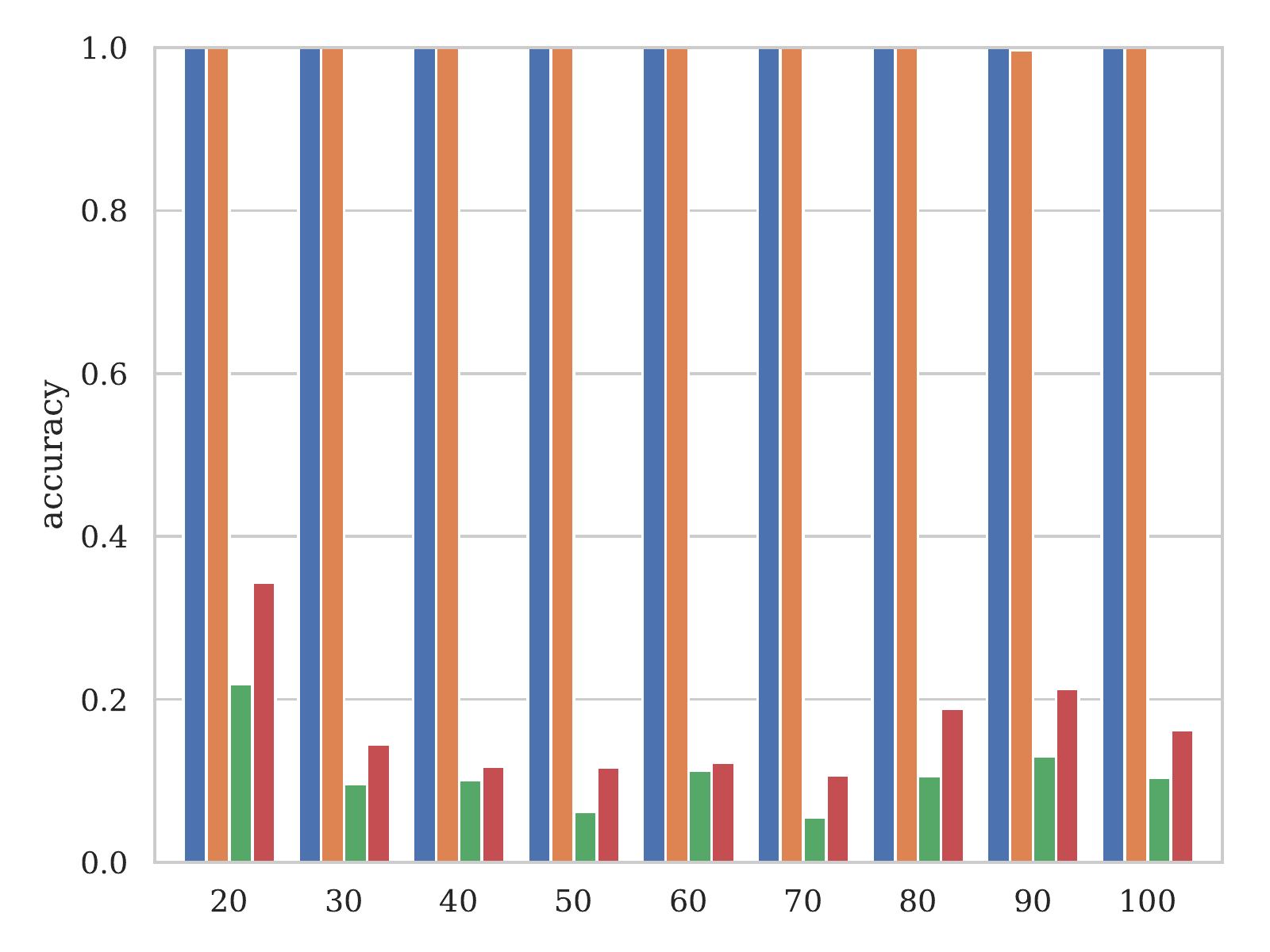}
	    \caption{\task{0}.}
	 \end{subfigure}
	 \begin{subfigure}{0.49\linewidth}
	    \includegraphics[width=\linewidth]{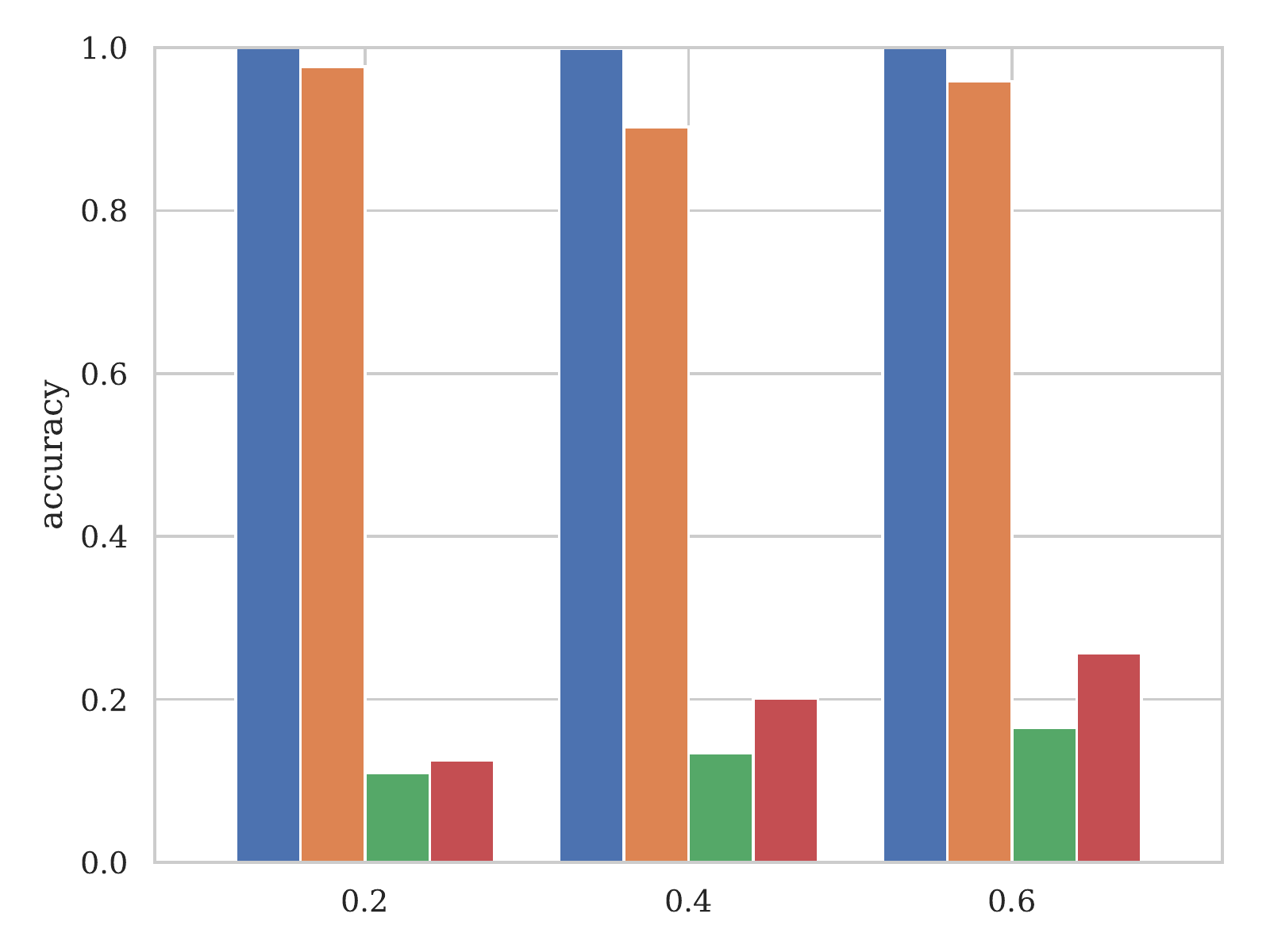}
	    \caption{\task{1}.}
	 \end{subfigure}
	 \begin{subfigure}{0.49\linewidth}
	    \includegraphics[width=\linewidth]{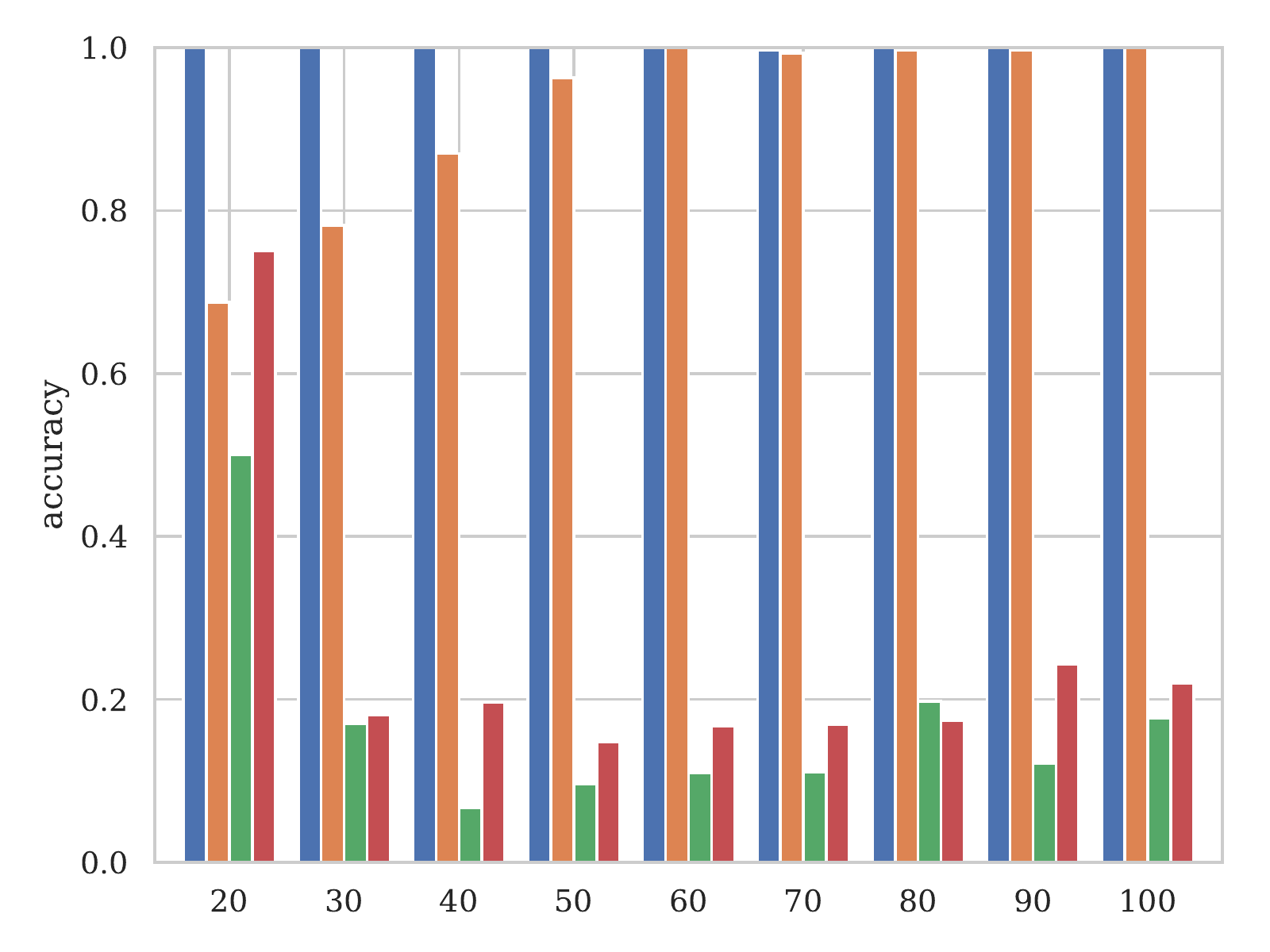} 
	    \caption{\task{1}.}
	 \end{subfigure}
	 
	 \begin{subfigure}{0.49\linewidth}
	    \includegraphics[width=\linewidth]{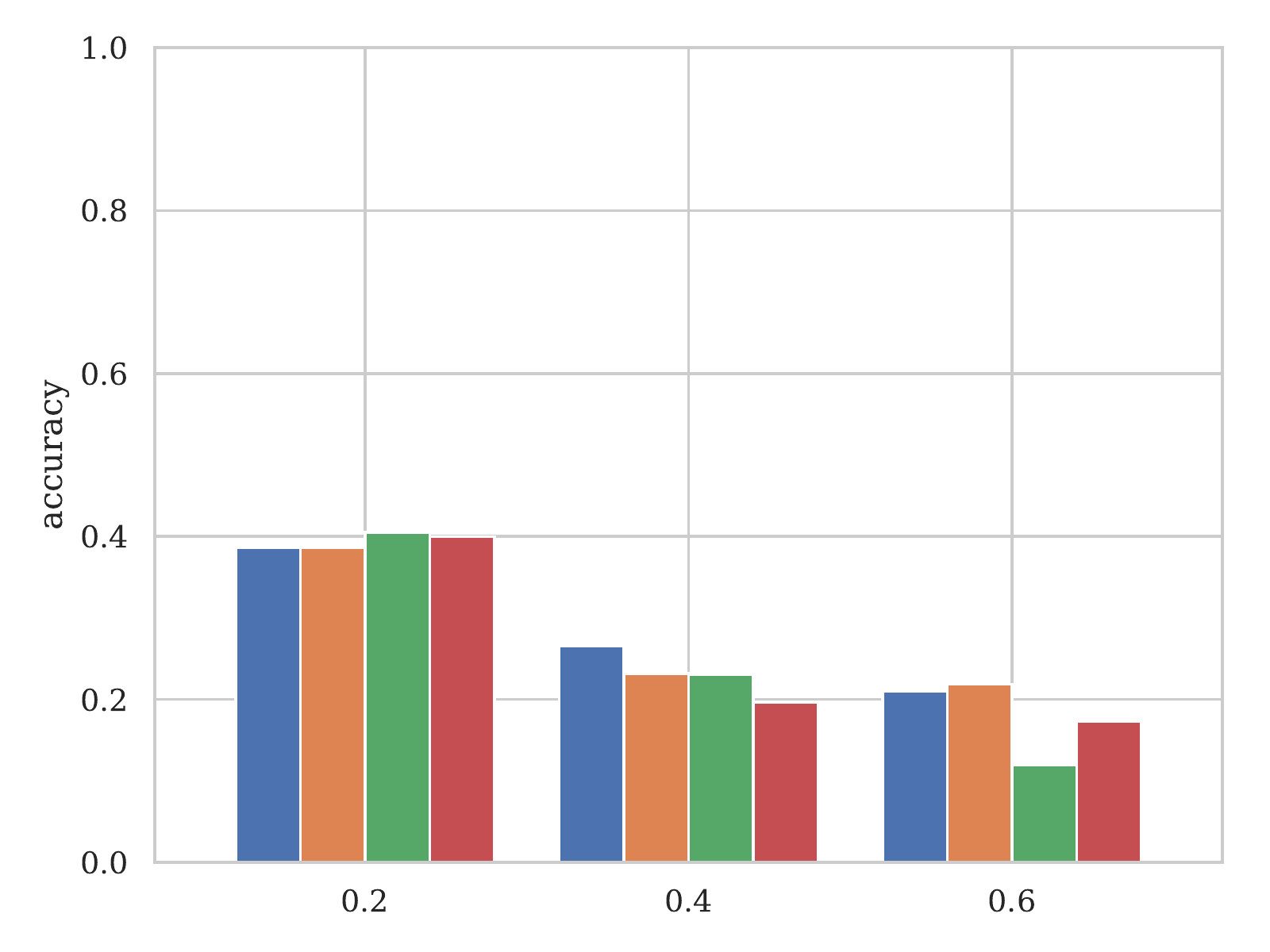}
	    \caption{\task{2}.}
	  \end{subfigure}
	  \begin{subfigure}{0.49\linewidth}
	    \includegraphics[width=\linewidth]{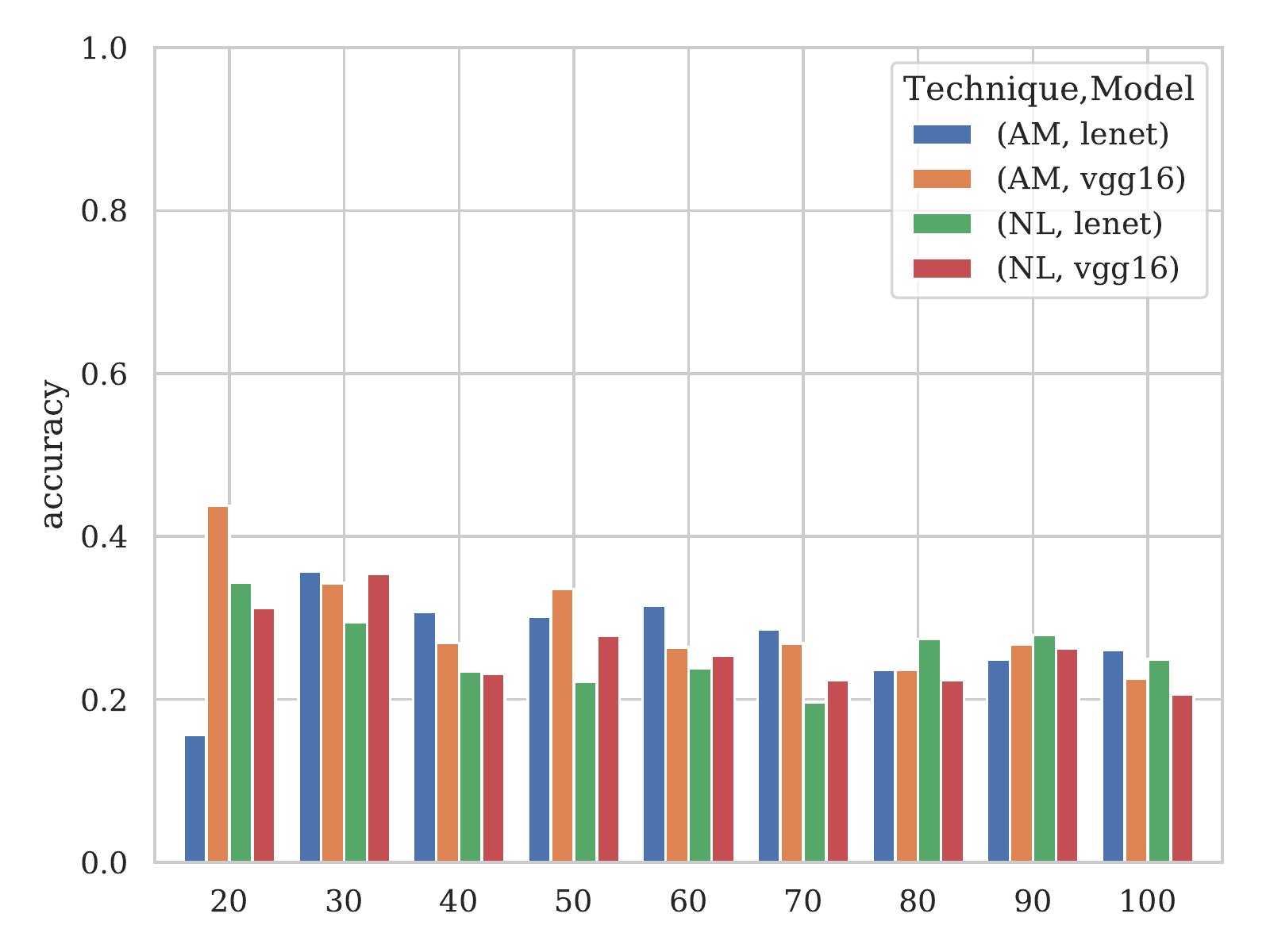} 
	    \caption{\task{2}.}
	 \end{subfigure}
	\caption{Accuracies on the test dataset for all the couples of network and representation grouped by range of edge densities (left column) and nodes (right column). The selected models maximize \accuracy on the validation dataset.\label{fig_perf_per_node_edge_accuracy}}
\end{figure}

Even if statistical tests validated our hypothesis, we have further analyzed the results to identify if one of the dataset generation parameters has an influence on the performances of the models.
Figure~\ref{fig_perf_per_node_edge_accuracy} groups the recognition performance per range of nodes and per density to locally identify the difficult cases on models that maximize the \accuracy metric.
\MD accuracy does not seem affected by either the nodes number or the edge density of the graph for \Task{0} and \Task{1}. \NL accuracy abruptly decreases when graph size exceeds \num{30} nodes and then remain stable up to \num{100} nodes. Moreover, \NL accuracy also seems to increase with increasing edge density which is surprising.	
However, \NL accuracies are so low that it may not be relevant to compare them.
Both \NL and \MD accuracies decrease when graph size or edge density increases for \Task{2}. Such an impact on the accuracy was expected and confirmed the results of~\cite{ghoniem2004comparison,ghoniem2005readability}.

\paragraph{\textbf{Models convergence}}
Another interesting aspect of such study is to determine which visualization technique is prone to fast convergence of the model.
We believe it is correlated to the difficulty of the problem: a fastest convergence means an easier problem and a slowest convergence means a more difficult problem. 
Table~\ref{tab_summary} shows that models trained on the \MD representation usually train faster (24 epochs to obtain the best model in average) than models trained with the \NL representation (39 epochs in average).
The difference is statistically significant as a Wilcoxon signed rank test computed on these 18 pairs obtained a p-value of 
$\num{0.016} << 0.05$%
.
Therefore, models learn faster how to solve the tasks with \MD representations than with \NL ones. 
Interpreting this learning speed is thorny as it could be only related to the time a network needs to learn the shape of the elements it is requested to count. This would not directly reflect the task difficulty itself. But even so, we believe that this kind of relation is induced by the visualization technique and can be interpreted.
Thus, it could be easier for users to learn how to solve the tasks on \MD representation than on \NL as models trained with \MD do converge faster. It would be interesting to investigate whether this assumption is relevant or not in a further study. 

\subsubsection{Limitations}
\begin{figure}[!bt]
	\centering
	\begin{subfigure}[b]{.4\linewidth}
		\includegraphics[width=\linewidth]{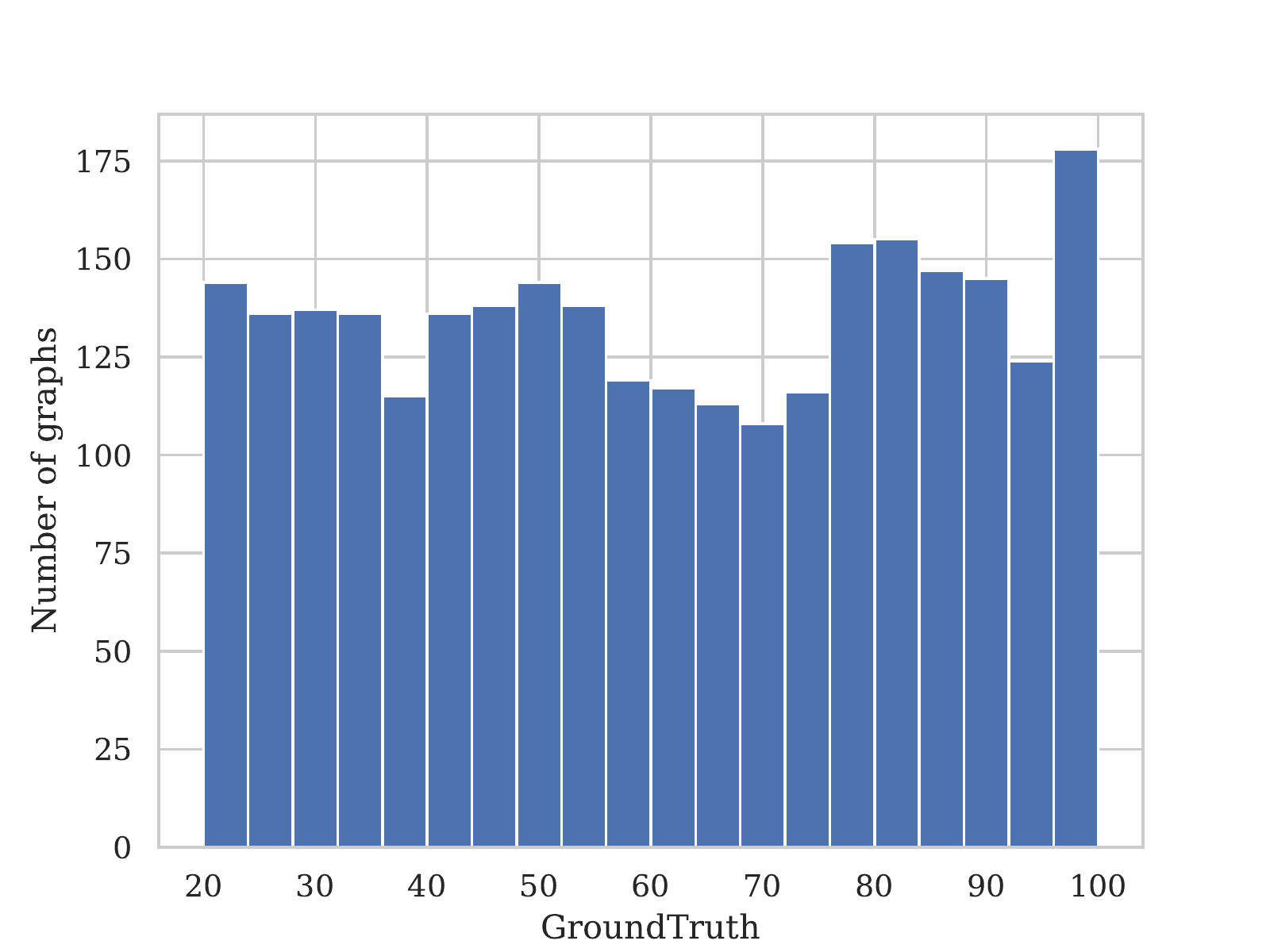}
		\caption{\Task{0}.}
	\end{subfigure}
	\begin{subfigure}[b]{.4\linewidth}
		\includegraphics[width=\linewidth]{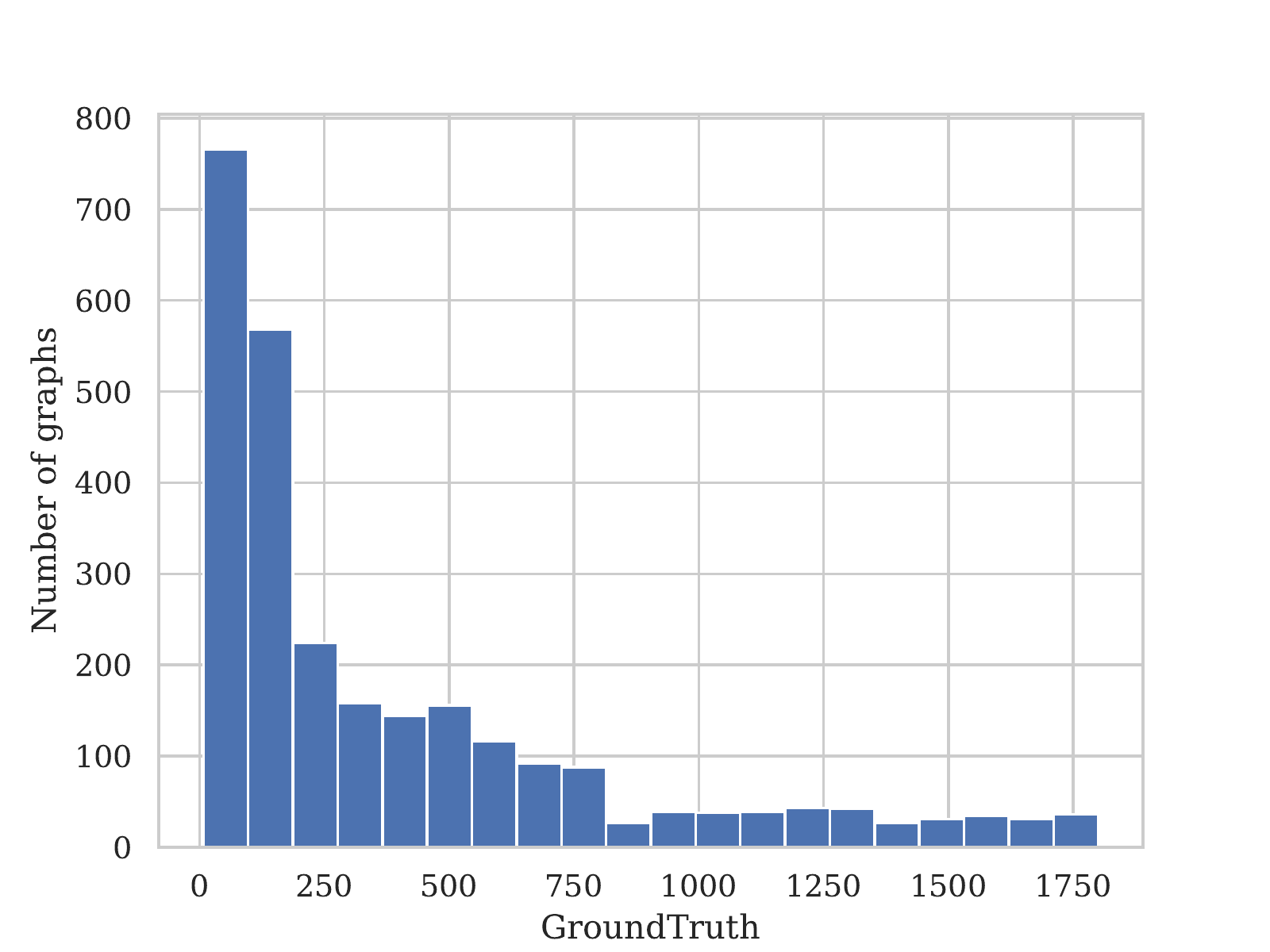}
		\caption{\Task{1}.}
	\end{subfigure}
	\begin{subfigure}[b]{.4\linewidth}
		\includegraphics[width=\linewidth]{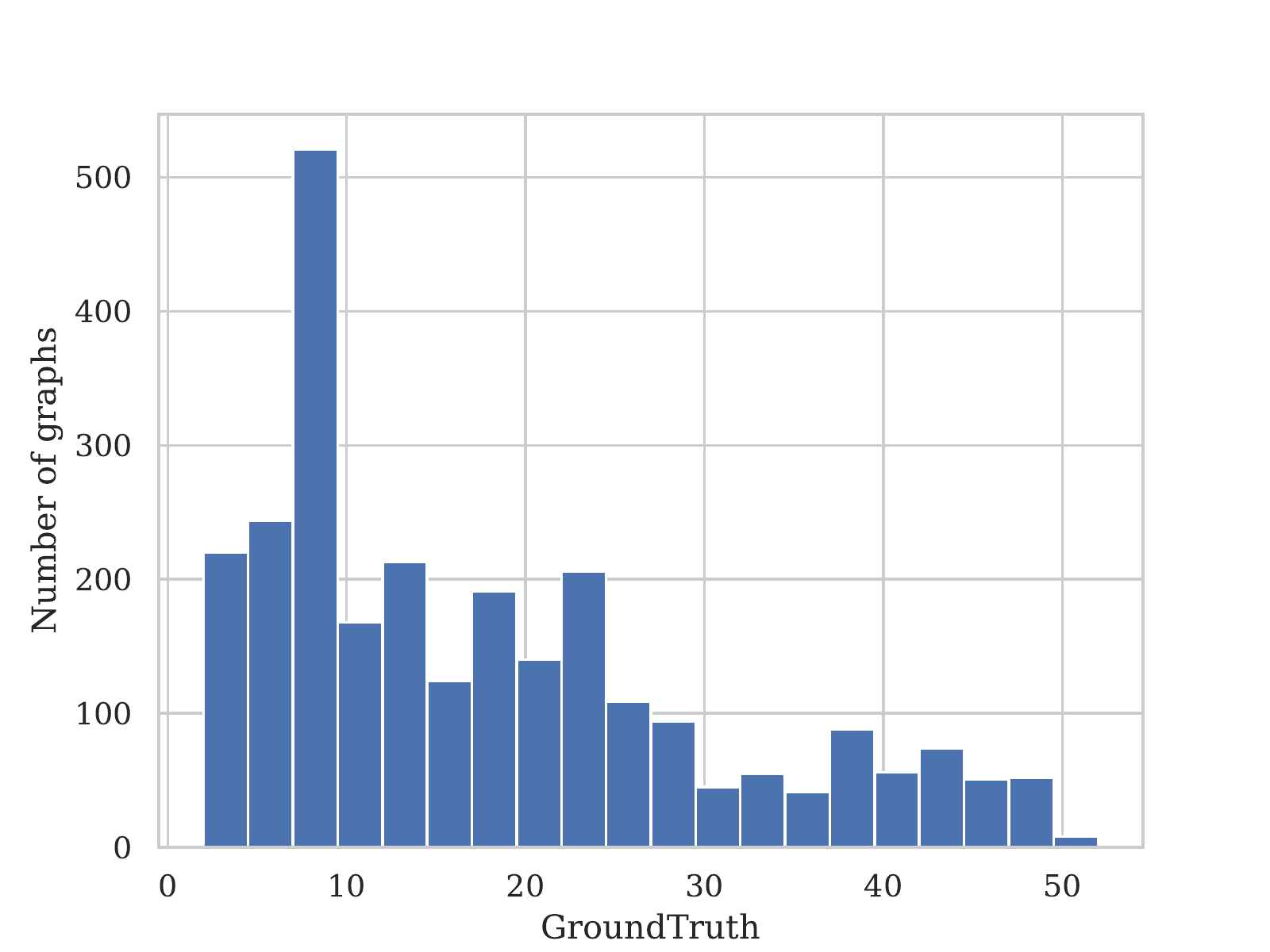}
		\caption{\Task{2}.}
	\end{subfigure}
	\begin{subfigure}[b]{.4\linewidth}
		\includegraphics[width=\linewidth]{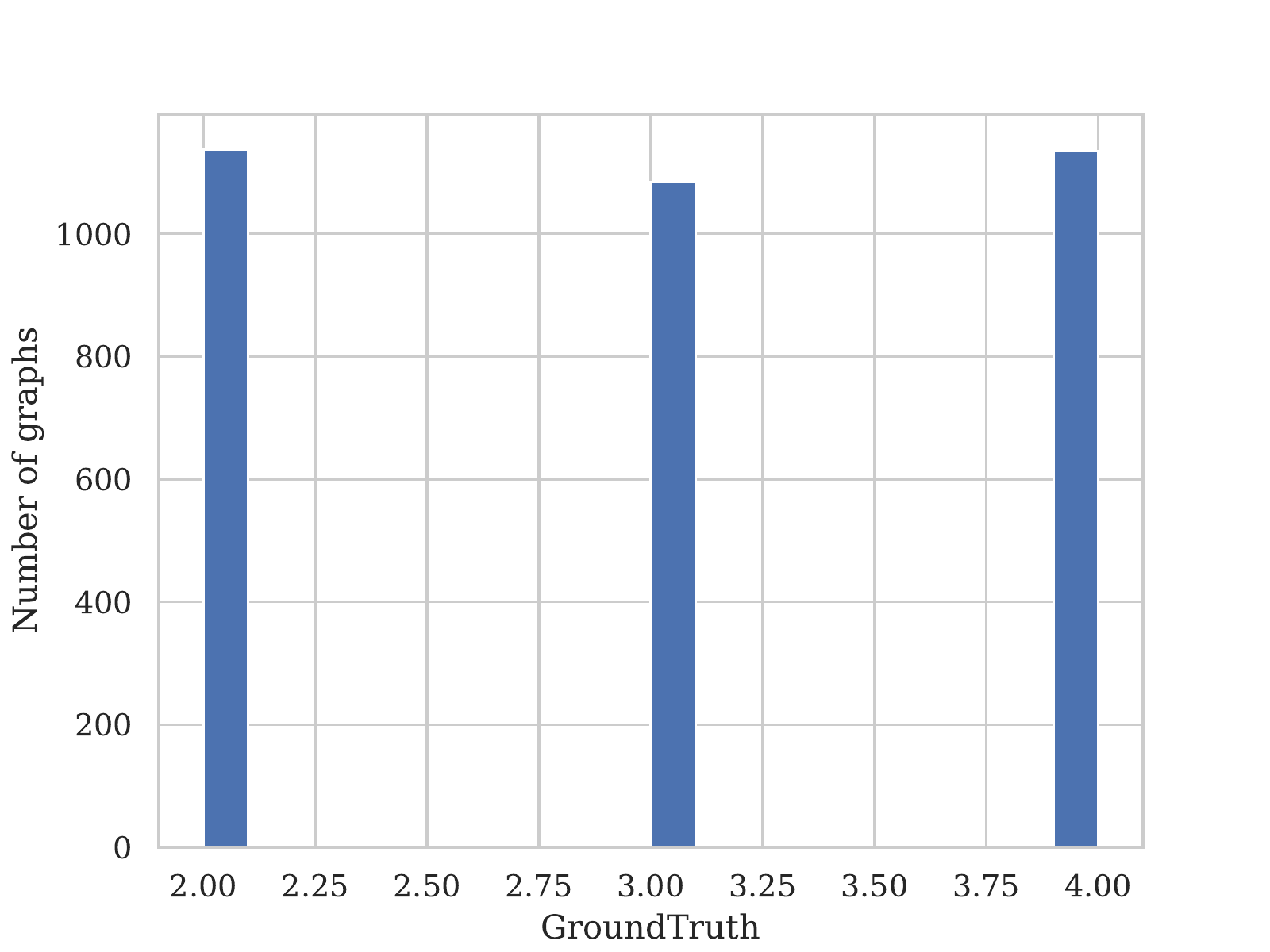}
		\caption{\Task{3}.}
	\end{subfigure}
	\caption{\label{fig_distribution}ground-truths distribution for the 4 tasks in the test database.}
\end{figure}

As for any experimental evaluation, the results of our use case can only be interpreted within the limits set by its evaluation protocol (datasets, algorithms used to generate layouts, resolution and visual choices to draw images, \emph{etc}).

Our database has been tailored for \task{0}; Figure~\ref{fig_distribution} presents the ground-truth distribution for the \num{3} tasks on the test database. As expected, the distribution of the number of nodes is nearly uniform (the variation is due to a sampling of the entire dataset). However, the distributions of the number of edges and of the maximum degree clearly do not follow a uniform distribution (but rather a power-law distribution) which was also expected. Indeed, small graphs with high edge densities may have the same number of edges than larger ones with low edge densities.
Even if it may have been better to generate a custom dataset for each task, the purpose of our study was to test whether the results of~\cite{ghoniem2004comparison, ghoniem2005readability} could be reproduced using computer vision techniques or not. As such unbalanced distributions are due to the dataset generation parameters from the original study, we could not avoid that bias.

To draw a graph with the \NL representation, it is first necessary to lay it out with a specific algorithm and its specific configuration parameters (Section~\ref{sec_representation}).
The models predictions, and the drawn conclusions, could be different depending on whether it is a non-force-directed algorithm or a force-directed algorithm.
We can assume CNN models may be more efficient if the graphs were drawn in an orthogonal manner (\textit{e.g.} \cite{Chimani04bend}).
Similar concerns arise in the \MD setting where it is known that several encodings can be used to draw the edges~\cite{sansen15} and several node orderings could be used to ease reading~\cite{behrisch2016matrix}.
However such limitations are inherent to any evaluation (automatic or human-based) and are not specifically related to our experiment.

The last main limitation we have identified lies in the image resolution and visual choices for its rendering. Concerning the \MD, our rendering engine renders grid-lines with a constant thickness (1 pixel), increasing the image resolution would increase rows height and columns width, it could also ease the completion of the considered counting tasks. One can have similar concerns about the rendering of \NL for which nodes size and edge thickness had to be arbitrarily set. Moreover, to reduce the computational time and memory consumption during the model training, we used grayscale images as mentioned in Section~\ref{sec_representation}. This drastically reduces the memory usage and the computational cost as the input image is coded with one channel instead of three in RGB. While color is usually considered as an important visual channel, we consider it should not penalize the models because the number of grayscale colors used is quite small (3) and are not ambiguous.
 
More generally, our \NL and \MD representations (\emph{i.e.} layout algorithm, image drawing, absence of interactions) were not the exact same as those from~\cite{ghoniem2004comparison, ghoniem2005readability} as their study do not provide enough details to reproduce entirely theirs.

\subsection{Shortest path length task of Okoe \textit{et al.}}

\subsubsection{Evaluation protocol}

\paragraph{\textbf{Tasks, visualization techniques and hypothesis}}

To study how the models perform on a task of higher level, we reproduced the \task{10} from the evaluation of ~\cite{okoe2018node}.
That task consists in determining the length of the shortest path between two highlighted nodes.
The evaluation showed that \NL performs better than \MD which was not contradictory with other studies evaluating that specific task (in particular with the results of \cite{ghoniem2004comparison, ghoniem2005readability}). 

In this paper, we refer to this task as \task{3}.
As for the first experiment, we expected our result to confirm the result of \cite{okoe2018node}:

\textit{H2:} \NL should outperform \MD for this task.

\paragraph{\textbf{Graphs dataset}} 
Unlike our first evaluation, the graphs we use for this experiment are those of the original study~\cite{okoe2018node}; they correspond to two real-world networks 
shared in a GitHub repository\footnote[1]{\url{github.com/mershack/Nodelink-vs-Adjacency-Matrices-Study}, last consulted on February 2020}, along with their study setup and results. 
In the original experiment setup, the answers to the \task{3} followed two rules: (i) there is always a path between the two highlighted nodes and (ii) the shortest path between these nodes is no more than 4. 
To generate as many task instances as possible (as it is required by deep learning training processes) while ensuring a uniform distribution of the ground-truths, we computed the number of shortest paths of length $2$ to $4$ for each graph. 
Considering the two graphs of the original setup, we could extract \num{5593} pairs of nodes for each graph and shortest path length ($2$, $3$ and $4$).
It led to \num{33558} tasks instances (\num{5593} pairs of nodes $*$ 3 path lengths $*$ 2 graphs).
This dataset is then randomly split into $3$ parts: $80\%$ for training set, $10\%$ for validation set and $10\%$ for test set.



\begin{figure*}[!tb]
	\centering

	 \begin{subfigure}{0.4\linewidth}
	    \includegraphics[width=\linewidth]{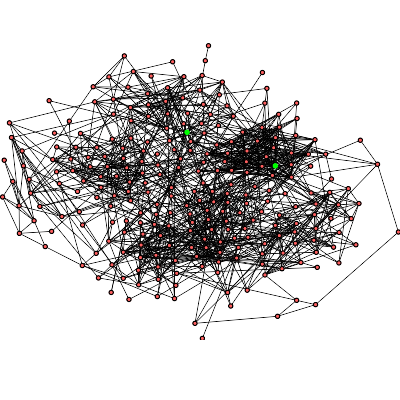}
		\caption{Receipe (\NL)}
	 \end{subfigure}
	 \begin{subfigure}{0.4\linewidth}
	    \includegraphics[width=\linewidth]{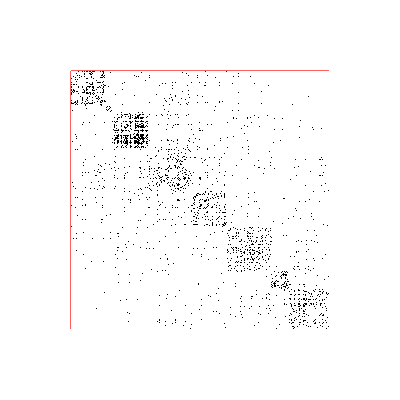}
		\caption{Receipe (\MD)}
	 \end{subfigure}
	 \begin{subfigure}{0.4\linewidth}
	    \includegraphics[width=\linewidth]{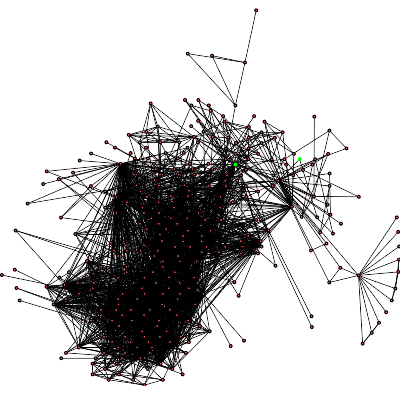}
		\caption{UsAir (\NL)}
	 \end{subfigure}
	 \begin{subfigure}{0.4\linewidth}
	    \includegraphics[width=\linewidth]{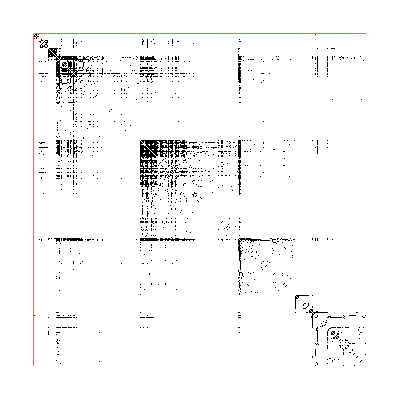}
		\caption{UsAir (\MD)}
	 \end{subfigure}
	\caption{\label{img_kob_dataset}Our \MD and \NL representations of the graphs of~\cite{okoe2018node}. Example of images of Receipe and UsAIr with \NL (a and b) and \ MD (b and d) where two nodes have been highlighted.}
\end{figure*} 

\paragraph{\textbf{Images datasets}}
\label{KobourovImageData}

In this experiment, we did not need to compute the graph layouts for the \NL representation and the node orderings for the \MD one, as \cite{okoe2018node} provide them in their GitHub repository. 

With the \MD representation used in the first experiment (see Section~\ref{GhoniemImagesData}), the generated images size would be about $2$ times the number of nodes in a graph, as every node shall be represented by at least $1$ pixel and a $1$ pixel grid line. 
For this experiment, such a representation would lead us to images of size $665 \times 665$ pixels which, coupled with the architecture of our models, is too large to feed them on our runtime infrastructure (see limitations in Section~\ref{KobourovLimitations}). 
To remedy this, we removed grid lines from the \MD representation and set the image size for both \NL and \MD representations to $400 \times 400$ pixels.
We generated RGB images where nodes were drawn in red (255,0,0) while nodes of interest were highlighted in green (0,255,0) (see Figure~\ref{img_kob_dataset}). 
These colors have been chosen so that (i) it is easy for a user to find the highlighted nodes, and (ii) the difference between these two colors is still significant once converted to grayscale as expected by the models. The second point is induced by our conversion formula which is: $L = 0.299 R + 0.587 G + 0.114 B$ where red and green are the two most important channels. 

\paragraph{\textbf{Deep convolutional networks}}
\label{KobourovDeepCNN}
As the approach we propose is based on the use of generic model architectures, the models we use, the number of epochs and the optimizer in this experiment are the same as in the first one (see Section~\ref{ghoniemDeepConv}).

Supposedly, the only changes to the architecture we would need to apply are in the input layer, to adapt the model input to our image size; and the output layer, to fit the number of classes the models need to predict in the current task. Therefore, the input of the models has been changed to a tensor of $400 \times 400$. 
As the maximum shortest path lengths (\emph{i.e.} diameters) in these graphs are $6$ and $7$, the number of classes to predict has been set to $8$.
However, even with these modifications, the models did not fit in the memory of our runtime infrastructure. Thus, we reduced the size of the fully connected layers of the two models hidden layers from $4096$ and $4096$ in the first experiment to $512$ and $256$. 
Finally, we needed to add to \vgg both a Convolution and a MaxPooling layer between the last convolution and the first fully connected layer in order to reduce memory usage when running our experiment.


\subsubsection{Results}


\paragraph{\textbf{Learning feasibility}}
Table~\ref{tab_learning} shows the $R^2$ scores of the best models for each network architecture and visualization.
One can see that except \lenet with \NL, none of the models were able to learn how to solve the task.
Indeed, \vgg with both \NL and \MD and \lenet with \MD all have $R^2$ scores lower than zero. 
By inspecting the predictions of these models, we noticed that at each epoch these models were only able to predict one value disregarding the input images.
In fact, unable to learn how to solve the task, the model learned the distribution of the dataset it has been trained on and does always predict the predominant value in the ground-truths so that it minimizes its error.
On the contrary, \lenet with \NL has a $R^2$ score of \num{0.46} which indicates that the model was able to learn how to solve \task{3}.



\paragraph{\textbf{Models performances and comparison}}


One can see in Table~\ref{tab_summary} that we only considered \accuracy and \mse metrics for \task{3}. Indeed, \percent and \deltametric would not have much meaning with the current models since they only predict $8$ values between $0$ and $7$. 
According to accuracy and MSE scores, \lenet obtained better results with \NL than with \MD. On the contrary, \vgg best models offer the same performance whatever the chosen metric and visualization. In fact, as these models always predict the same value and as the ground-truths only contains $3$ values evenly distributed, their accuracies are about $33\%$.
We followed the same procedure as in the first experiment to assess whether a model outperforms another or not. First, we selected the best model for each metric, then we computed the mean absolute difference between the ground-truths and the models predictions and finally, we compared these absolute differences using a statistical (Wilcoxon) test.
Table~\ref{tab_md_best_nl} shows that the mean absolute difference between \MD and \NL is significant with \lenet architecture as the p-value is $<<0.05$. 
It shows that, considering these two model architectures, \task{3} is solved more accurately with \NL representations than with \MD representation.

\paragraph{\textbf{Models convergence}} 
With \lenet architecture, the fact that the model trained on \MD representation has not been able to learn in this experiment has major impacts on our results interpretation. 
We may consider that because the model did not learn, it is impossible to compare its convergence speeds to others.
Nevertheless, if the convergence speed were to be analyzed, we consider that if a model was not able to learn how to solve the task, it may be due to the bound on number of epochs. 
For instance, \lenet on \MD might have been able to learn to solve the task with a hundred more training epochs. In that case, \lenet with \NL outperforms \lenet with \MD as its best results are reached at epoch \num{76}.
Though, we rather believe that if a model cannot learn a task on a representation, it either means that the representation prevents to solve the task, or that the representation is too complex to solve such a task. 

As \vgg models were unable to learn with both \MD and \NL, there is no convergence to take into account. We will discuss some possible reasons of that failures in Section~\ref{sec_discussion}.


We can conclude that our hypothesis \textit{H2} has been verified as there is a significant performance difference in favor of \NL representation between selected models (see Table~\ref{tab_md_best_nl}). Yet, this result is to be considered within the scope of that experiment.

\subsubsection{Limitations}
\label{KobourovLimitations}

The first limitation concerns memory issues  faced by our runtime infrastructure when feeding large images to the networks. We had two runtimes environments at our disposal to train them: (i) a GPU computer with a NVIDIA Titan X (Pascal) 12GB card,  and (ii) a CPU cluster of 50 nodes (\emph{i.e.} computers) with 64GB of RAM each among which up to 45GB could be reserved for training execution. We achieved to train \lenet on the 12GB GPU computer with some modifications (see Section~\ref{KobourovDeepCNN}), but this memory was not enough to run \vgg. For this last, we had to run on the CPU cluster, which took about $6$ days to train over $100$ epochs. We needed to scale down our architectures to run this experiment, although it goes against our will to make use of generic implementations of models architectures. We expect such drawback will be present in any other study treating high-level tasks or large images. 

Like in the  first experiment, the remaining limitations are related to the graph drawings. First, we still ignored any interaction; this is even more limiting as the graphs are few hundreds nodes big and we could not scale up the image size as it would have been necessary to keep the nodes drawings wide enough. Thus, nodes in our representations are significantly small ($1$ pixel per node in \MD images) and interactions such as zoom-in or selection, like in the original study, bring a lot to a visualization technique in these cases. Nevertheless, we want to compare the visualizations techniques themselves, not the interactive tools they are brought with.

Finally, our way of highlighting a node (see Section~\ref{KobourovImageData}) differs from the original study. We defined a fair way to do it so that it is possible for a human and a learning model to identify them. The main issue is that since nodes are represented with only $1$ pixel in the \MD representation, it could be tough for a network to understand which nodes are highlighted (mainly because of MaxPooling layers). This difficulty comes from our infrastructure's technical limitations. But even so, such a limitation is at first induced by the \MD representation itself, which will always occupy a quadratic space on the nodes number.

\section{Discussion}
\label{sec_discussion}

While it seems that users and performance of computer vision techniques can be correlated, our approach also suffers from several limitations. This section discusses the main identified limitations of our approach.

The user is usually asked to solve a given task in an interactive environment in most of the formal user evaluations.
For example, in the studies of \cite{ghoniem2004comparison,ghoniem2005readability}, the user can zoom in/out, pan or even highlight the neighborhood of a given node to ease task completion. On one hand, providing interaction tools may make the evaluation more realistic as no visualization software would provide only still images. On the other hand, it may restrict the scope of the evaluation as it would not only compare several visualization techniques but rather combinations of visualizations and their associated interaction tools. This is particularly true when comparing multiform visualization techniques where the provided interaction tools have to be adapted to each technique.

Although our experiments covered different task types and difficulties, some further work shall be made to strengthen this study. Indeed, these experiments are not sufficient to conclude that users and  performance of computer vision techniques can be generally correlated. Comparing more diverse tasks, on more visualization techniques would be a first step to generalize our study and explore further how correlated the results of users and models can be.

Another drawback of our approach is related to its computational cost. Performing the proposed evaluation was more difficult than expected. \vgg and even \lenet can be heavy to run on some infrastructure when the input shape becomes too large.
For the first experiment, and with the distributed infrastructure we described in Section~\ref{KobourovLimitations}, we could train all the models with a batch size of \num{400} images in one week.
For \vgg, train a model (100 epochs) for one task and one visualization took one day, against 4 hours for \lenet. 
For the second experiment we had to downsize our models and batch sizes during training so they could fit in memory. With the previously described settings, each \vgg took about 6 days to train over 100 epochs for each visualization technique, against 9 hours for \lenet. Such a computation time is a limitation of our approach. We assume that progress in machine learning dedicated hardware will soon enable to run such experiments in less than a day. Nevertheless, today's infrastructures are sufficient to deepen this exploration.

The second experiment raised another concern about the way we designed our study. The way we normalize the models training parameters causes problems during the training of some of them. It is expected that some models will be better design for some tasks than others. But as previously stated, we want our different models to be trained in a way that the visualization technique (\emph{i.e.} the features they are provided) are the only difference that can impact their performances. However, the second experiment showed that it might be more complicated than expected as the different model's architectures we use and their inherent complexity make it difficult to set a generalist environment that makes them all learn properly. 
In that experiment, we noticed that the accuracies of the \vgg models across the epochs oscillated. Such a behavior could be due to a too high learning rate that makes the gradient descent steps too large to optimize the cost function; or a lack of normalization. 
Other explanations could be hypothetized, for instance, \vgg might have been unable to learn due to a lack of training samples, a too short training period, or even a too large batch size.

Finally, it is noteworthy to mention that the model which best learned to solve \task{3} converged in about \num{80} epochs and \num{100} epochs were not enough with others, whereas the mean of convergence speed for the 3 counting tasks is \num{31} epochs. Thus, it could be interesting to study whether this result means that there is a correlation between the task's difficulty and the model convergence speed or if they are only due to some other parameter.

\section{Conclusion}
\label{sec_conclusion}
This paper has presented some experiments that explore correlations between humans and computer vision techniques based on deep learning; to evaluate graph visualization techniques. 
We trained models to solve tasks on several representations (\MD and \NL) and compared their performances to two state-of-the-art user evaluations.
Our results tend to show that humans and computer vision techniques could be correlated on the considered tasks types, however additional work is needed to verify its generalization.

A first improvement to deepen the experiment we ran is to enlarge its scope with more tasks of various types and use other models architectures to verify our results robustness. A second lead of improvement is to compare a representation with itself while only changing some of its parameters, to make sure that our models are learning to interpret the visualization. For example, one could compare two identical \MD representations which use a state of the art and a random ordering algorithm and verify that our approach gives an expected result \cite{mueller2007comparison}: random ordering performs worse.

The purpose of this study was to test the feasibility of our approach. As its results draw positive feedback, we hope it could serve as a first step for further studies around this thematic.

A direction for future research is to include interactive visualization in the evaluation process. Reinforcement learning~\cite{mnih2013playing} seems to be a good option as with such technique an agent could interact with the visualization to learn how to solve a task.
A second direction could explore further the (dis)similarities between humans and machines computations. One could study the correlation between machine performance metrics (\emph{e.g.} convergence speed) and humans performances, or how classes of problems affect humans and machines performances and if such changes are correlated.

Finally, if future studies were to verify our assumption, we hope that this new approach could lead toward a new methodology that uses computer vision techniques as a pre-process of user evaluations. We believe that such a methodology could bring a lot to the visualization community by providing it a tool to make experiments reproducible, create test beds and improve actual end-user evaluation's protocol definition.

\bibliographystyle{plain}
\bibliography{ia4vis}

\end{document}